\documentclass[aps,prd,twocolumn,superscriptaddress]{revtex4}
\usepackage{epsfig,epsf}
\usepackage{amsmath}
\usepackage{amsthm}
\usepackage{amsfonts}
\usepackage{amssymb}
\usepackage{dsfont}
\usepackage{multirow}
\usepackage{appendix}
\usepackage{slashed}
\usepackage[active]{srcltx}
\usepackage{psfrag}

\begin{document}

\title{Testing the doubly charged charm-strange tetraquarks}
\date{\today}
\author{S.~S.~Agaev}
\affiliation{Department of Physics, Kocaeli University, 41380 Izmit, Turkey}
\affiliation{Institute for Physical Problems, Baku State University, Az--1148 Baku,
Azerbaijan}
\author{K.~Azizi}
\affiliation{Department of Physics, Do\v{g}u\c{s} University, Acibadem-Kadik\"{o}y, 34722
Istanbul, Turkey}
\affiliation{School of Physics, Institute for Research in Fundamental Sciences (IPM),
P.~O.~Box 19395-5531, Tehran, Iran}
\author{H.~Sundu}
\affiliation{Department of Physics, Kocaeli University, 41380 Izmit, Turkey}

\begin{abstract}
The spectroscopic parameters and decay channels of the doubly charged
scalar, pseudoscalar and axial-vector charm-strange tetraquarks $Z_{%
\overline{c}s}=[sd][\overline u \overline c]$ are explored within framework
of the QCD sum rule. The masses and current couplings of these
diquark-antidiquark states are calculated by means of two-point correlation
functions and taking into account the vacuum condensates up to eight
dimensions. To compute the strong couplings of $Z_{\overline{c}s}$ states
with $D,\ D_{s},\ D^{\ast},\ D_{s}^{\ast},\ D_{s1}(2460),\
D_{s0}^{\ast}(2317),\ \pi$ and $K$ mesons we use QCD light-cone sum rules
and evaluate width of their $S$- and $P$-wave decays to a pair of negatively
charged conventional mesons: For the scalar state $Z_{\overline{c}s}\to D_s
\pi,\ DK, \ D_{s1}(2460)\pi$, for the pseudoscalar state $Z_{\overline{c}s}
\to D_{s}^{\ast}\pi,\ D^{\ast}K, \ D_{s0}^{\ast}(2317)\pi,$ and for the
axial-vector state $Z_{\overline{c}s} \to D_{s}^{\ast}\pi,\ D^{\ast}K,\
D_{s1}(2460)\pi$ decays are investigated. Obtained predictions for the
spectroscopic parameters and decay widths of the $Z_{\overline{c}s}$
tetraquarks may be useful for experimental investigations of the doubly
charged exotic hadrons.
\end{abstract}

\maketitle

\section{Introduction}

During last decade tetraquarks, i.e. bound states of four quarks are in the
center of intensive experimental and theoretical investigations. Starting
from discovery of the famous resonance $X(3872)$ in $B$ meson decay $%
B\rightarrow KX\rightarrow KJ/\psi \rho \rightarrow KJ/\psi \pi ^{+}\pi ^{-}$
by Belle \cite{Choi:2003ue}, and after observation of the same state by
other groups \cite{Abazov:2004kp,Acosta:2003zx,Aubert:2004ns} experimental
collaborations collected valuable information on the spectroscopic
parameters and decay channels of the exotic states. They were discovered in
various inclusive and exclusive hadronic processes. In this connection it is
worth to note $B$ meson decays, $e^{+}e^{-}$ and $\overline{p}p$
annihilations and $pp$ collisions. Theoretical studies of exotic hadrons,
apart from tetraquark states, include pentaquarks and hybrid mesons and
encompass variety of models and calculational methods claiming to explain
the internal structure of these states and calculate their experimentally
measured parameters. Comprehensive information on collected experimental
data and detailed analysis of theoretical achievements and existing problems
can be found in latest review works Refs.\ \cite%
{Chen:2016qju,Chen:2016spr,Esposito:2016nozN,Esposito:2014rxa,Meyer:2015eta}.

The great success in physics of the exotic hadrons is connected with
discovery of charged multiquark resonances. The first charged tetraquarks,
namely $Z^{\pm}(4430)$ states were observed by the Belle Collaboration in $B$
meson decays $B \to K\psi^{\prime} \pi^{\pm}$ as resonances in the $%
\psi^{\prime}\pi^{\pm}$ invariant mass distributions \cite{Choi:2007wga}.
The resonances $Z^{+}(4430)$ and $Z^{-}(4430)$ were detected and studied by
Belle in the processes $B \to K\psi^{\prime} \pi^{+}$ \cite{Mizuk:2009da}
and $B^{0} \to K^{+}\psi^{\prime} \pi^{-}$ \cite{Chilikin:2013tch}, as well.
These states constitute an important subclass of multiquark systems, because
charged resonances can not be explained as excited charmonium or bottomonium
states, and therefore, are real candidates to genuine tetraquarks.

Hadrons built of four quarks of different flavors form another intriguing
class in the tetraquark family. Depending on a quark content these states
may be neutral or charged particles. Among the observed tetraquarks the $%
X(5568)$ resonance remains a unique candidate to a hadron composed of four
different quarks. At the same time it is a particle containing $b$-quark,
i.e. is an open bottom tetraquark. The evidence for $X(5568)$ was first
reported by the D0 Collaboration in Ref.\ \cite{D0:2016mwd}. Later it was
observed again by D0 in the $B_{s}^{0}$ meson's semileptonic decays \cite{D0}%
. But other experimental groups, namely the LHCb and CMS collaborations
could not find this resonance from analysis of their experimental data \cite%
{Aaij:2016iev,CMS:2016}, which make the experimental situation around $%
X(5568)$ unclear and controversial. Numerous theoretical works devoted to
investigation of $X(5568)$ resonance's structure and calculation of its
parameters led also to contradictory conclusions. The results of these
studies are in a reasonable agreement with measurements carried out by the
D0 Collaboration, while in other works an existence of the $X(5568)$ state
is an object of discussions \cite%
{Agaev:2016mjb,Chen:2016mqt,Wang:2016mee,Wang:2016tsi,Zanetti:2016wjn,Agaev:2016ijz,Dias:2016dme,Wang:2016wkj,Xiao:2016mho,Agaev:2016urs,Burns:2016gvy,Guo:2016nhb,Lu:2016zhe,Albaladejo:2016eps,Lang:2016jpk,Esposito:2016itg,Kang:2016zmv}%
. The detailed analysis of problems related to the status of the $X(5568)$
resonance can be found in original papers (see for instance, Ref.\ \cite%
{Chen:2016spr} and references therein).

The  tetraquarks which might carry double electric charge  constitute another interesting
class of exotic hadrons \cite{Esposito:2013fma}. These hypothetical particles if observed
can be interpreted as diquark-antidiquark states: Formation of molecular states from two 
mesons of same charge is almost impossible due to repulsive forces between them. 
The doubly charged particles may exist, for example, as double charmed  tetraquarks
$[cc][\bar d \bar s]$ or $[cc][\bar s \bar s]$. In other words, they may contain
two or three quark flavors. The phenomenology of these states, their decay modes and production
mechanisms were investigated in Ref.\ \cite{Esposito:2013fma}. In the context of the lattice QCD
the mass spectra of these particles were evaluated in the paper  \cite{Guerrieri:2014nxa}.
As it was revealing recently  the tetraquarks containing  quarks
of four different flavors may also carry double electric charge \cite{Chen:2017rhl}.
In fact, it is
not difficult to see that tetraquarks $Z_{\overline{c}s}=[sd][\overline u
\overline c]$ and $Z_{c \overline{s}}= [uc][\overline {s} \overline {d}]$
belong to this category of particles, and at the same time, are open charm
states. Authors of Ref.\ \cite{Chen:2017rhl} wrote down also possible $S$-
and $P$-wave decay channels of these states. Strictly speaking, the open
charm tetraquarks were previously investigated in the literature (see, for
example Refs. \ \cite{Galkin,Agaev:2016lkl,Liu:2016ogz}). The spectroscopic
parameters and decay widths of the open charm tetraquark containing three
different light quarks were calculated in Ref.\ \cite{Agaev:2016lkl}. In
this study the open charm tetraquark was considered as a partner of the $%
X(5568)$ state. In other words, the quark content of $X_c=[su][\overline{c}%
\overline{d}]$ was obtained from $X_b=[su][\overline{b}\overline{d}]$ by $b
\to c$ replacement. Due to differences in the charges of $b$ and $c$ quarks
the partner state $X_c$ does not bear the same charge as $X_b$. This
conclusion is true in the case of $Z_{\overline{c}s}$, as well. If the state
$Z_{\overline{c}s}$ bears the charge $-2|e|$, its $b$-partner $Z_{\overline{b%
}s}$ has $-|e|$. In general, there do not exist doubly charged tetraquarks
composed of $b$ and three different light quarks. The genuine doubly charged
tetraquarks with $b$ belong to a subclass of open charm-bottom particles and
should contain also $c$-quark. For example, the state $Z_{b\overline{c}%
}=[bs][\overline{u}\overline{c}]$ has the charge $-2|e|$.

In the present work we are going to concentrate on features of doubly
charged charm-strange tetraquarks $Z_{\overline{c}s}$ with spin-parity $%
J^{P}=0^{+},\ 0^{-}$ and $1^{+}$, and calculate their masses, current
couplings and decay widths. To this end, we use QCD two-point sum rule
approach by including into analysis quark, gluon and mixed vacuum
condensates up to eight dimensions, and evaluate their spectroscopic
parameters. Obtained results are employed to reveal kinematically allowed
decay channels of the tetraquarks $Z_{\overline{c}s}$. They also enter as
input parameters to expressions of the corresponding decay widths. We
calculate the width of decay channels $Z_{\overline{c}s}\rightarrow
D_{s}\pi,\ DK$, and $D_{s1}(2460)\pi$ (for $J^{P}=0^{+}$), $Z_{\overline{c}%
s}\rightarrow D_{s}^{\ast }\pi ,\ D^{\ast }K$ and $D_{s0}^{\ast}(2317)\pi$
(for $J^{P}=0^{-}$), as well as $Z_{\overline{c}s}\rightarrow D_{s}^{\ast
}\pi ,\ D^{\ast }K$ and $D_{s1}(2460)\pi $ (in the case of $J^{P}=1^{+}$).
For these purposes, we analyze vertices of the tetraquarks $Z_{\overline{c}%
s} $ with the conventional mesons, and evaluate the corresponding strong
couplings using QCD sum rules on the light-cone. The QCD light-cone sum rule
method is one of the powerful nonperturbative tools to explore parameters of
the conventional hadrons \cite{Balitsky:1989ry}. In the case of vertices
built of a tetraquark and two conventional mesons the standard methods of
the light-cone sum rules should be supplemented by a technique of an
approach known as the "soft-meson" approximation \cite%
{Belyaev:1994zk,Ioffe:1983ju}. For investigation of the exotic states the
light-cone sum rules method was adapted in Ref. \cite{Agaev:2016dev}, and
successfully applied for analysis of various tetraquarks' decays \cite%
{Agaev:2016dsg,Agaev:2017uky,Agaev:2017foq,Agaev:2017tzv}.

This article is organized in the following manner. In Sec.\ \ref{sec:Masses}
we calculate the masses and current couplings of the doubly charged scalar,
pseudoscalar and axial-vector charm-strange tetraquarks $Z_{\overline{c}%
s}=[sd][\overline{u}\overline{c}]$ by treating them as diquark-antidiquark
systems. In Sec.\ \ref{sec:ScalDec} we consider the decays of the doubly
charged scalar tetraquark to $D_{s}\pi $, $DK$ and $D_{s1}(2460)\pi $ final
states. The Section \ref{sec:PsVDec} is devoted to decay channels of the
pseudoscalar and axial-vector tetraquarks. Here we compute width of their
decays to $D_{s}^{\ast }\pi$, $D^{\ast }K$ and $D_{s0}^{\ast}(2317)\pi $
(for $0^{-}$ state), and to $D_{s}^{\ast }\pi$, $D^{\ast }K$ and $%
D_{s1}(2460)\pi $ (for $1^{+}$ state). Section \ref{sec:Concl} is reserved
for our concluding remarks.


\section{Spectroscopic parameters of the scalar, pseudoscalar and axial
vector tetraquarks $Z_{\overline{c}s}$}

\label{sec:Masses}
In this section we calculate the mass and current coupling of the $Z_{%
\overline{c}s}=[sd][\overline{u}\overline{c}]$ tetraquarks with the quantum
numbers $J^{P}=0^{+}$, $0^{-}$ and $1^{+}$ by treating them as
diquak-antidiquark systems. In order to simplify the expressions we
introduce the notations: in what follows the scalar tetraquark $Z_{\overline{%
c}s}$ will be denoted as $Z_{S}$, whereas for the pseudoscalar and
axial-vector ones we will utilize $Z_{PS}$ and $Z_{AV}$, respectively.

The scalar tetraquarks within the context of two-point sum rule approach can
be explored using interpolating currents of $C\gamma _{5}\otimes \gamma
_{5}C $ or $C\gamma _{\mu }\otimes \gamma ^{\mu }C$ types, where $C$ is the
charge conjugation operator. In the present work we restrict ourselves by
the simplest case and employ the current%
\begin{equation}
J=s_{a}^{T}C\gamma _{5}d_{b}\left( \overline{u}_{a}\gamma _{5}C\overline{c}%
_{b}^{T}-\overline{u}_{b}\gamma _{5}C\overline{c}_{a}^{T}\right) .
\label{eq:Curr1}
\end{equation}%
To study the pseudoscalar and axial-vector tetraquarks $Z_{PS}$ and $Z_{AV%
\text{ }}$ we utilize $C\gamma _{\mu }\otimes \gamma _{5}C$ type
interpolating current
\begin{equation}
J_{\mu }=s_{a}^{T}C\gamma _{\mu }d_{b}\left( \overline{u}_{a}\gamma _{5}C%
\overline{c}_{b}^{T}-\overline{u}_{b}\gamma _{5}C\overline{c}_{a}^{T}\right)
.  \label{eq:Curr2}
\end{equation}%
Then the correlation functions $\Pi (p)$ and $\Pi _{\mu \nu }(p)$ necessary
for the sum rule computations take the forms
\begin{equation}
\Pi (p)=i\int d^{4}xe^{ipx}\langle 0|\mathcal{T}\{J(x)J^{\dagger
}(0)\}|0\rangle ,  \label{eq:CF1}
\end{equation}%
and
\begin{equation}
\Pi _{\mu \nu }(p)=i\int d^{4}xe^{ipx}\langle 0|\mathcal{T}\{J_{\mu
}(x)J_{\nu }^{\dagger }(0)\}|0\rangle .  \label{eq:CF2}
\end{equation}%
The current $J_{\mu }$ couples to both the pseudoscalar $0^{-}$ and
axial-vector $1^{+}$ states, therefore the function $\Pi _{\mu \nu }(p)$ can
be used to calculate parameters of the $Z_{PS}$ and $Z_{AV\text{ }}$
tetraquarks.

We start our analysis from calculations of the scalar state's spectroscopic
parameters. In accordance with QCD sum rule method the correlator given by
Eq.\ (\ref{eq:CF1}) should be expressed in terms of physical parameters of
the $Z_{S}$ state. In the case under analysis $\Pi (p)$ takes simple form
and is defined by the equality
\begin{equation}
\Pi ^{\mathrm{Phys}}(p)=\frac{\langle 0|J|Z_{S}(p)\rangle \langle
Z_{S}(p)|J^{\dagger }|0\rangle }{m_{Z_{S}}^{2}-p^{2}}+...,  \label{eq:Phys1}
\end{equation}%
where $m_{Z_{S}}$ is the mass of the $Z_{S}$ state, and dots stand for
contributions of the higher resonances and continuum states. In order to
simplify $\Pi ^{\mathrm{Phys}}(p)$ we introduce the matrix element
\begin{equation}
\langle 0|J|Z_{S}(p)\rangle =f_{Z_{S}}m_{Z_{S}},
\end{equation}%
where $m_{Z_{S}}$ and $f_{Z_{S}}$ are the mass and current coupling of $%
Z_{S}(p)$. Then, for the correlation function we obtain
\begin{equation}
\Pi ^{\mathrm{Phys}}(p)=\frac{m_{Z_{S}}^{2}f_{Z_{S}}^{2}}{m_{Z_{S}}^{2}-p^{2}%
}+\ldots .  \label{eq:Phys1A}
\end{equation}%
The Borel transformation applied to $\Pi ^{\mathrm{Phys}}(p)$ yields
\begin{equation}
\mathcal{B}\Pi ^{\mathrm{Phys}%
}(p)=m_{Z_{S}}^{2}f_{Z_{S}}^{2}e^{-m_{Z_{S}}^{2}/M^{2}}+\ldots ,
\label{eq:Phys1AB}
\end{equation}%
where $M^{2}$ is the Borel parameter.

The same correlation function $\Pi (p)$ calculated in terms of the
quark-gluon degrees of freedom reads%
\begin{eqnarray}
&&\Pi ^{\mathrm{QCD}}(p)=i\int d^{4}xe^{ipx}\left\{ \mathrm{Tr}\left[ \gamma
_{5}\widetilde{S}_{c}^{b^{\prime }b}(-x)\gamma _{5}S_{u}^{a^{\prime }a}(-x)%
\right] \right.  \notag \\
&&\times \mathrm{Tr}\left[ S_{s}^{aa^{\prime }}(x)\gamma _{5}\widetilde{S}%
_{d}^{bb^{\prime }}(x)\gamma _{5}\right] -\mathrm{Tr}\left[ \gamma _{5}%
\widetilde{S}_{c}^{a^{\prime }b}(-x)\right.  \notag \\
&&\left. \times \gamma _{5}S_{u}^{b^{\prime }a}(-x)\right] \mathrm{Tr}\left[
S_{s}^{aa^{\prime }}(x)\gamma _{5}\widetilde{S}_{d}^{bb^{\prime }}(x)\gamma
_{5}\right]  \notag \\
&&-\mathrm{Tr}\left[ \gamma _{5}\widetilde{S}_{c}^{b^{\prime }a}(-x)\gamma
_{5}S_{u}^{a^{\prime }b}(-x)\right] \mathrm{Tr}\left[ S_{s}^{aa^{\prime
}}(x)\gamma _{5}\widetilde{S}_{d}^{bb^{\prime }}(x)\gamma _{5}\right]  \notag
\\
&&\left. +\mathrm{Tr}\left[ \gamma _{5}\widetilde{S}_{c}^{a^{\prime
}a}(-x)\gamma _{5}S_{u}^{b^{\prime }b}(-x)\right] \mathrm{Tr}\left[
S_{s}^{aa^{\prime }}(x)\gamma _{5}\widetilde{S}_{d}^{b^{\prime }b}(x)\gamma
_{5}\right] \right\} .  \notag \\
&&{}  \label{eq:QCD1}
\end{eqnarray}%
Here we use the short-hand notation%
\begin{equation}
\widetilde{S}_{q(c)}^{ab}(x)=CS_{q(c)}^{Tab}(x)C,  \label{eq:Not}
\end{equation}%
with $S_{q}(x)$ and $\ S_{c}(x)$ being the $q=u,\ d$ and $c$-quark
propagators, respectively.

The QCD sum rules to evaluate $m_{Z_{S}}$ and $f_{Z_{S}}$ can be obtained by
choosing the same Lorentz structures in both of $\Pi ^{\mathrm{Phys}}(p)$
and $\Pi ^{\mathrm{QCD}}(p)$, and equating the relevant invariant
amplitudes. \ In the case under investigation the only Lorentz structure
which exists in $\Pi (p)$ is one $\sim I$. For calculation of the mass and
coupling it is convenient to employ the two-point spectral density $\rho
_{0}^{\mathrm{QCD}}(s)$. In terms of $\rho _{0}^{\mathrm{QCD}}(s)$ the
invariant amplitude $\Pi ^{\mathrm{QCD}}(p^{2})$ can be written down as the
dispersion integral
\begin{equation}
\Pi ^{\mathrm{QCD}}(p^{2})=\int_{(m_{c}+m_{s})^{2}}^{\infty }\frac{\rho
_{0}^{\mathrm{QCD}}(s)}{s-p^{2}}ds+....
\end{equation}%
By applying the Borel transformation to $\Pi ^{\mathrm{QCD}}(p^{2})$ ,
equating the obtained expression with $\mathcal{B}\Pi ^{\mathrm{Phys}%
}(p^{2}) $, and subtracting the contribution due to higher excited and
continuum states we find the final sum rules. For the mass of the $Z_{S}$
state it is given by the formula
\begin{equation}
m_{Z_{S}}^{2}=\frac{\int_{(m_{c}+m_{s})^{2}}^{s_{0}}dss\rho _{0}^{\mathrm{QCD%
}}(s)e^{-s/M^{2}}}{\int_{(m_{c}+m_{s})^{2}}^{s_{0}}ds\rho _{0}^{\mathrm{QCD}%
}(s)e^{-s/M^{2}}},  \label{eq:SRmass}
\end{equation}%
whereas for the current coupling $f_{Z_{S}}$ we get
\begin{equation}
f_{Z_{S}}^{2}=\frac{1}{m_{Z_{S}}^{2}}%
\int_{(m_{c}+m_{s})^{2}}^{s_{0}}dse^{(m_{Z_{S}}^{2}-s)/M^{2}}\rho _{0}^{%
\mathrm{QCD}}(s).  \label{eq:SRc}
\end{equation}%
In Eqs.\ (\ref{eq:SRmass}) and (\ref{eq:SRc}) $s_{0}$ is the continuum
threshold parameter which separates contributions stemming from the
ground-state and ones due to higher resonances and continuum states. The $%
M^{2}$ and $s_{0}$ are two important auxiliary parameters of sum rule
computations choices of which should meet some requirements which will be
shortly explained below.

In the case of the current $J_{\mu }$ the correlation function $\Pi _{\mu
\nu }^{\mathrm{Phys}}(p)$ derived using the physical parameters of
tetraquarks contains two terms. In fact, the current $J_{\mu }$ couples to
the pseudoscalar and axial-vector tetraquarks, therefore after inserting
into Eq. \ (\ref{eq:CF2}) full set of states and integrating over $x$ we get
expression containing contributions of the ground state pseudoscalar and
axial-vector particles, i. e.
\begin{eqnarray}
&&\Pi _{\mu \nu }^{\mathrm{Phys}}(p)=\frac{\langle 0|J_{\mu
}|Z_{AV}(p)\rangle \langle Z_{AV}(p)|J_{\nu }^{\dagger }|0\rangle }{%
m_{Z_{AV}}^{2}-p^{2}}  \notag \\
&&+\frac{\langle 0|J_{\mu }|Z_{PS}(p)\rangle \langle Z_{PS}(p)|J_{\nu
}^{\dagger }|0\rangle }{m_{Z_{PS}}^{2}-p^{2}}+...,  \label{eq:Phys2}
\end{eqnarray}%
with $m_{Z_{PS}}$ and $m_{Z_{AV}}$ being the mass of the pseudoscalar and
axial-vector states, respectively. Here again by dots we denote
contributions coming from higher excitations and continuum states in both
the pseudoscalar and vector channels. In general one may consider only one
of these terms, and compute parameters of the chosen particle with $%
J^{P}=0^{-}$ or $1^{+}$. In the present work we are interested in both of
these particles, therefore keep explicitly two terms in $\Pi _{\mu \nu }^{%
\mathrm{Phys}}(p)$, and use different structures to derive two sets of sum
rules.

Further simplification of\ $\Pi _{\mu \nu }^{\mathrm{Phys}}(p)$ can be
achieved by expressing the relevant matrix elements in terms of the masses $%
m_{Z_{PS}}$, $m_{Z_{AV}}$ and current couplings $f_{Z_{AV}}$ , $f_{Z_{PS}}$
\begin{eqnarray*}
\langle 0|J_{\mu }|Z_{AV}(p)\rangle &=&f_{Z_{AV}}m_{Z_{AV}}\varepsilon _{\mu
}(p), \\
\langle 0|J_{\mu }|Z_{PS}(p)\rangle &=&f_{Z_{PS}}m_{Z_{PS}}p_{\mu }.
\end{eqnarray*}%
Then it is easy to show that
\begin{eqnarray}
&&\Pi _{\mu \nu }^{\mathrm{Phys}}(p)=\frac{m_{Z_{AV}}^{2}f_{Z_{AV}}^{2}}{%
m_{Z_{AV}}^{2}-p^{2}}\left( -g_{\mu \nu }+\frac{p_{\mu }p_{\nu }}{p^{2}}%
\right)  \notag \\
&&+\frac{m_{Z_{PS}}^{2}f_{Z_{PS}}^{2}}{m_{Z_{PS}}^{2}-p^{2}}p_{\mu }p_{\nu
}+...  \label{eq:Phys2A}
\end{eqnarray}%
In order to obtain the function $\Pi _{\mu \nu }^{\mathrm{QCD}}(p)$ we
substitute the interpolating current $J_{\mu }$\ from Eq.\ (\ref{eq:Curr2})
into Eq.\ (\ref{eq:CF2}), and contract the quark fields. As a result, for $%
\Pi _{\mu \nu }^{\mathrm{QCD}}(p)$ we get:
\begin{eqnarray}
&&\Pi _{\mu \nu }^{\mathrm{QCD}}(p)=i\int d^{4}xe^{ipx}\left\{ \mathrm{Tr}%
\left[ \gamma _{5}\widetilde{S}_{c}^{a^{\prime }b}(-x)\gamma
_{5}S_{u}^{b^{\prime }a}(-x)\right] \right.  \notag \\
&&\times \mathrm{Tr}\left[ S_{s}^{aa^{\prime }}(x)\gamma _{\nu }\widetilde{S}%
_{d}^{bb^{\prime }}(x)\gamma _{\mu }\right] +\mathrm{Tr}\left[ \gamma _{5}%
\widetilde{S}_{c}^{b^{\prime }a}(-x)\right.  \notag \\
&&\left. \times \gamma _{5}S_{u}^{a^{\prime }b}(-x)\right] \mathrm{Tr}\left[
S_{s}^{aa^{\prime }}(x)\gamma _{\nu }\widetilde{S}_{d}^{bb^{\prime
}}(x)\gamma _{\mu }\right]  \notag \\
&&-\mathrm{Tr}\left[ \gamma _{5}\widetilde{S}_{c}^{b^{\prime }b}(-x)\gamma
_{5}S_{u}^{a^{\prime }a}(-x)\right] \mathrm{Tr}\left[ S_{s}^{aa^{\prime
}}(x)\gamma _{\nu }\widetilde{S}_{d}^{bb^{\prime }}(x)\gamma _{\mu }\right]
\notag \\
&&\left. -\mathrm{Tr}\left[ \gamma _{5}\widetilde{S}_{c}^{a^{\prime
}a}(-x)\gamma _{5}S_{u}^{b^{\prime }b}(-x)\right] \mathrm{Tr}\left[
S_{s}^{aa^{\prime }}(x)\gamma _{\nu }\widetilde{S}_{d}^{bb^{\prime
}}(x)\gamma _{\mu }\right] \right\} .  \notag \\
&&{}  \label{eq:CF3}
\end{eqnarray}%
The correlation function $\Pi _{\mu \nu }^{\mathrm{QCD}}(p)$ has the
following Lorentz structures
\begin{eqnarray}
&&\Pi _{\mu \nu }^{\mathrm{QCD}}(p)=\Pi _{AV}^{\mathrm{QCD}}(p^{2})\left(
-g_{\mu \nu }+\frac{p_{\mu }p_{\nu }}{p^{2}}\right)  \notag \\
&&+\frac{p_{\mu }p_{\nu }}{p^{2}}\Pi _{PS}^{\mathrm{QCD}}(p^{2}),
\label{eq:QCD2}
\end{eqnarray}%
where $\Pi _{AV}^{\mathrm{QCD}}(p^{2})$ and $\Pi _{PS}^{\mathrm{QCD}}(p^{2})$
are invariant amplitudes corresponding to the axial-vector and pseudoscalar
tetraquarks, respectively. Equating the structures $\sim g_{\mu \nu }$ in
Eqs.\ (\ref{eq:Phys2A}) and (\ref{eq:QCD2}), and performing the Borel
transformation it is possible we derive the sum rules for parameters of the
axial-vector tetraquark: They are given by Eqs.\ (\ref{eq:SRmass}) and (\ref%
{eq:SRc}) but with $\rho _{0}^{\mathrm{QCD}}(s)$ replaced by $\rho _{V}^{%
\mathrm{QCD}}(s)$.

The sum rules for the pseudoscalar state are found by computing $p^{\mu }\Pi
_{\mu \nu }^{\mathrm{Phys}}(p)$ and $p^{\mu }\Pi _{\mu \nu }^{\mathrm{QCD}%
}(p),$ and \ matching obtained expressions, which consist of terms with
parameters of the pseudoscalar tetraquark. Then for the mass of the
pseudoscalar state we again find the sum rule \ (\ref{eq:SRmass}), but $\rho
_{0}^{\mathrm{QCD}}(s)\rightarrow \rho _{S}^{\mathrm{QCD}}(s)$, whereas the
coupling $f_{Z_{PS}}$ is determined by the following expression
\begin{equation}
f_{Z_{PS}}^{2}=\frac{1}{m_{Z_{PS}}^{4}}%
\int_{(m_{c}+m_{s})^{2}}^{s_{0}}dse^{(m_{Z_{PS}}^{2}-s)/M^{2}}\rho _{S}^{%
\mathrm{QCD}}(s).  \label{eq:SRc2}
\end{equation}%
In the present work we calculate the two-point spectral densities $\rho
_{0}^{\mathrm{QCD}}(s),$ $\rho _{V}^{\mathrm{QCD}}(s)$ and $\rho _{S}^{%
\mathrm{QCD}}(s)$ by taking into account quark, gluon and mixed vacuum
condensates up to eight dimensions.

The sum rules (\ref{eq:SRmass}), (\ref{eq:SRc}) and (\ref{eq:SRc2}) depend
on the masses of $c$ and $s$-quarks, and vacuum expectations of quark, gluon
and mixed operators, which are presented below:
\begin{eqnarray}
&&m_{c}=(1.27\pm 0.03)~\mathrm{GeV},\ \ m_{s}=96_{-4}^{+8}~\mathrm{MeV}
\notag \\
&&\langle \bar{q}q\rangle =-(0.24\pm 0.01)^{3}\ \mathrm{GeV}^{3},\ \langle
\bar{s}s\rangle =0.8\ \langle \bar{q}q\rangle ,  \notag \\
&&m_{0}^{2}=(0.8\pm 0.1)\ \mathrm{GeV}^{2},\ \langle \overline{q}g_{s}\sigma
Gq\rangle =m_{0}^{2}\langle \overline{q}q\rangle ,  \notag \\
&&\langle \overline{s}g_{s}\sigma Gs\rangle =m_{0}^{2}\langle \bar{s}%
s\rangle ,  \notag \\
&&\langle \frac{\alpha _{s}G^{2}}{\pi }\rangle =(0.012\pm 0.004)\,\mathrm{GeV%
}^{4},  \notag \\
&&\langle g_{s}^{3}G^{3}\rangle =(0.57\pm 0.29)\ \mathrm{GeV}^{6}.
\label{eq:Param}
\end{eqnarray}%
For condensates we use their standard values, whereas the masses of the
quarks are borrowed from Ref.\ \cite{Olive:2016xmw}. In the chiral limit
adopted in the present work $m_{u}=m_{d}=0$.

The sum rules contain also, as it has been just noted above, the auxiliary
parameters $M^{2}$ and $s_{0}$. It is clear, that physical quantities
evaluated from the sum rules should not depend on the Borel parameter and
continuum threshold, but in real calculations one can only reduce their
effect to a minimum. In fixing of working regions for $M^{2}$ and $s_{0}$
some conditions should be obeyed. Thus, we fix the upper bound $M_{\mathrm{%
max}}^{2}$ of the window $M^{2}\in \lbrack M_{\mathrm{min}}^{2},\ M_{\mathrm{%
max}}^{2}]$ for the Borel parameter by requiring fulfilment of the following
constraint
\begin{equation}
\mathrm{PC}=\frac{\Pi ^{\mathrm{QCD}}(M_{\mathrm{max}}^{2},\ s_{0})}{\Pi ^{%
\mathrm{QCD}}(M_{\mathrm{max}}^{2},\ \infty )}>0.13,  \label{eq:Rest1}
\end{equation}%
where $\Pi ^{\mathrm{QCD}}(M^{2},\ s_{0})=\mathcal{B}\Pi ^{\mathrm{QCD}%
}(p^{2})$ is the Borel transform of the invariant amplitude after the
continuum subtraction. Minimal limit for $\mathrm{PC}$ \ chosen as $\sim 0.1$
is smaller than in the case of the conventional mesons, but is typical for
multiquark systems. The lower limit of the same region $M_{\mathrm{min}}^{2}$
is deduced from convergence of the operator product expansion. By
quantifying this condition we require that contribution of the last term in
OPE should not exceed $5\%$, i. e.
\begin{equation}
\frac{\Pi ^{\mathrm{QCD}(\mathrm{Dim}8)}(M_{\mathrm{min}}^{2},\ \infty )}{%
\Pi ^{\mathrm{QCD}}(M_{\mathrm{min}}^{2},\ \infty )}<0.05,  \label{eq:Rest2}
\end{equation}%
has to be obeyed. Another condition for the lower limit is exceeding of the
perturbative contribution the nonperturbative one. In the present work we
apply the following criterion: at the lower bound of $M^{2}$ the
perturbative contribution has to constitute $\geq 60\%$ part of the full
result.

\begin{figure}[h]
\begin{center}
\includegraphics[totalheight=6cm,width=8cm]{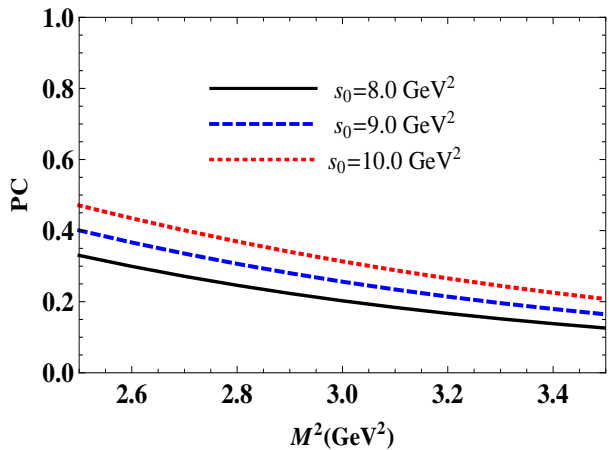}\,\,
\end{center}
\caption{ The pole contribution in the case of the scalar tetraquark vs the
Borel parameter $M^{2}$ at different $s_{0}$.}
\label{fig:PCS}
\end{figure}

Analysis of the sum rules for the $Z_{S}$ state enable us to fix the Borel
and continuum threshold parameters within the limits:
\begin{equation}
M^{2}\in \lbrack 2.5-3.5]~\mathrm{GeV}^{2},\ s_{0}~\in \lbrack 8-10]\
\mathrm{GeV}^{2}.  \label{eq:BTparam}
\end{equation}%
In these regions the pole contribution defined by Eq.\ (\ref{eq:Rest1}) is $%
\mathrm{PC}>0.14$. At the same time, contribution coming from the pole term
at $M_{\mathrm{min}}^{2}$ constitutes $\sim 65\%$, and at $M_{\mathrm{max}%
}^{2}$ approximately $60\%$ of the sum rule (\ref{eq:SRmass}) used to
evaluate the mass of $Z_{S}$ state. The convergence of OPE expansion in
these regions is also satisfied. Thus, contribution of the Dim8 term in OPE
does not exceed $3\%$ . All these features are seen in Figs.\ \ref{fig:PCS}, %
\ref{fig:ConvS} and \ref{fig:Pert.NpertS} , where we plot the pole
contribution, contributions due to different nonperturbative terms, and the
perturbative and total nonperturbative components of $\Pi ^{\mathrm{QCD}%
}(M^{2},\ s_{0})$ to demonstrate that in the regions for $M^{2}$ and $s_{0}$
given by Eq.\ (\ref{eq:BTparam}) constraints imposed on $\Pi ^{\mathrm{QCD}%
}(M^{2},\ s_{0})$ are fulfilled.

\begin{widetext}

\begin{figure}[h!]
\begin{center}
\includegraphics[totalheight=6cm,width=8cm]{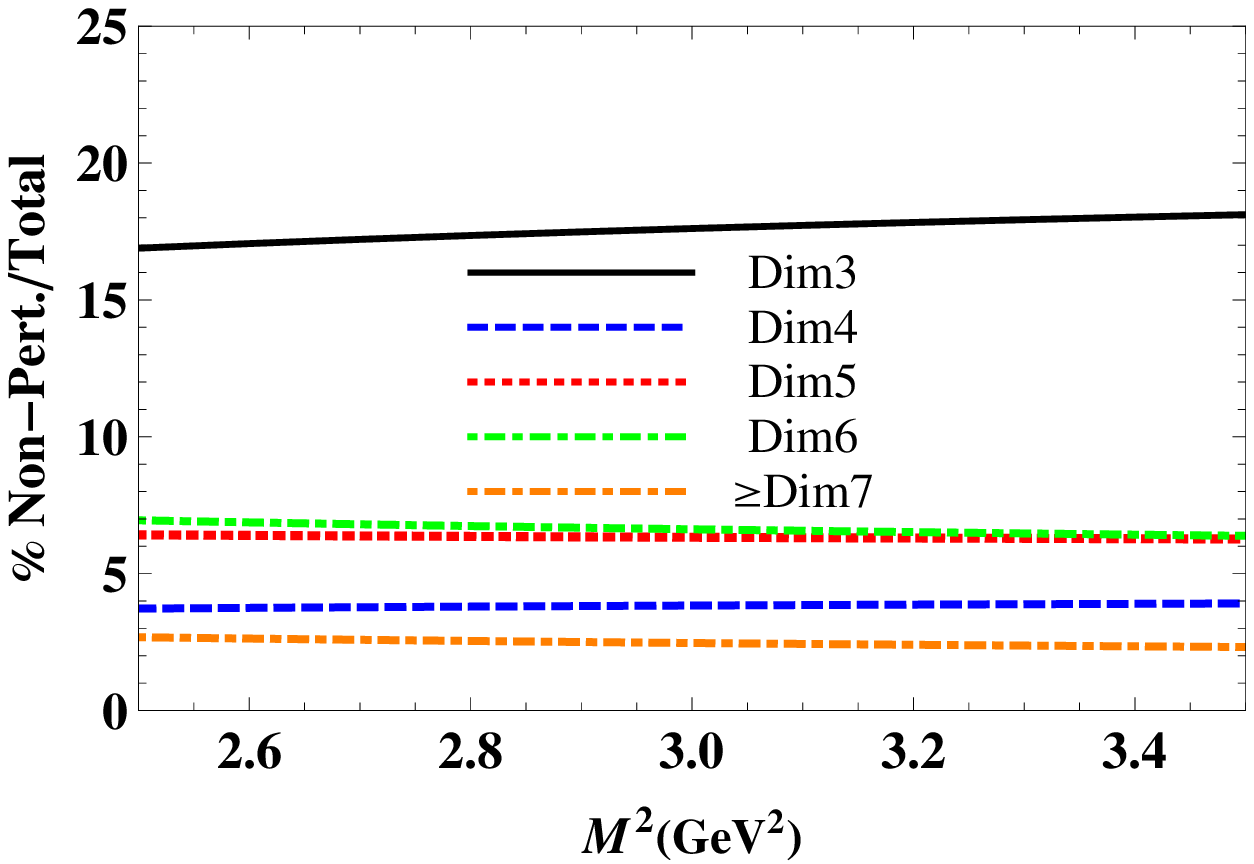}\,\, %
\includegraphics[totalheight=6cm,width=8cm]{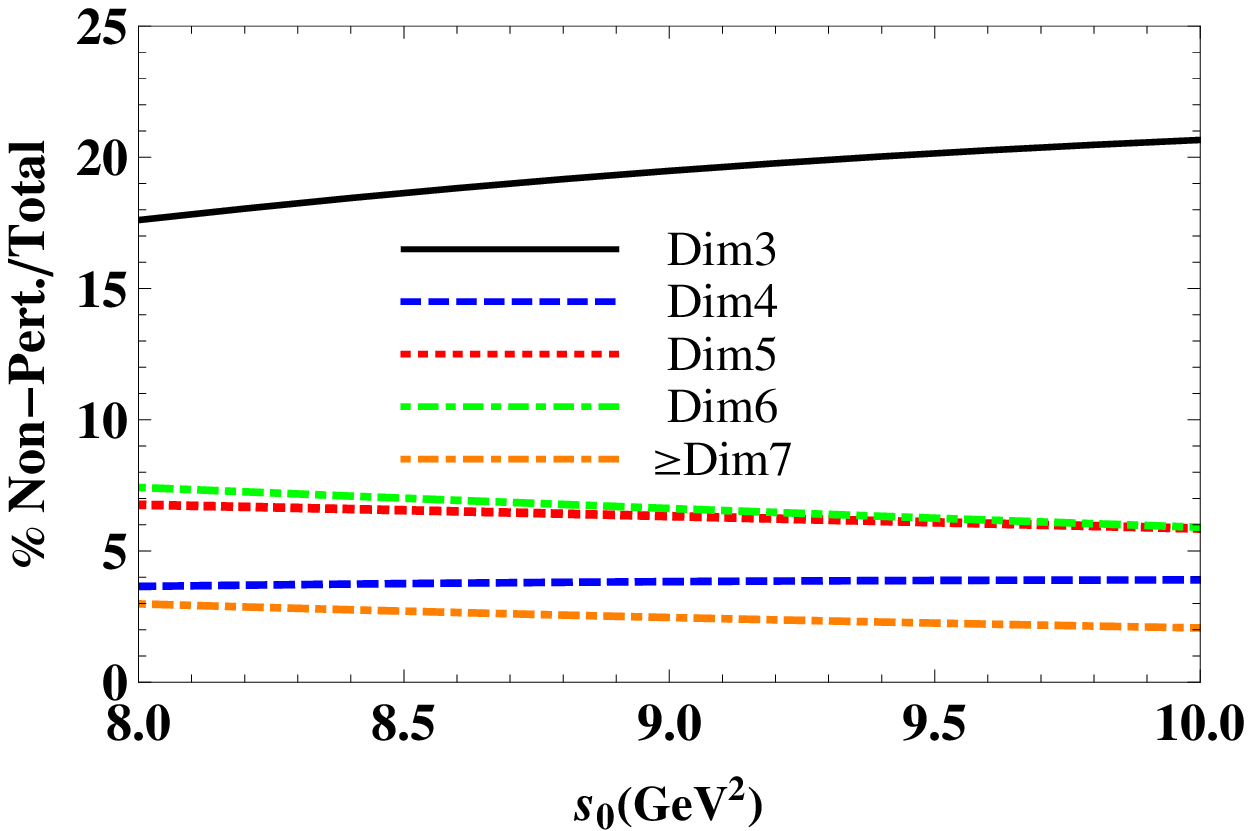}
\end{center}
\caption{  Contributions due to different  nonperturbative terms in the case of $Z_S$ as functions of $M^2$ (left panel), and $s_0$ (right
panel). } \label{fig:ConvS}
\end{figure}

\begin{figure}[h!]
\begin{center}
\includegraphics[totalheight=6cm,width=8cm]{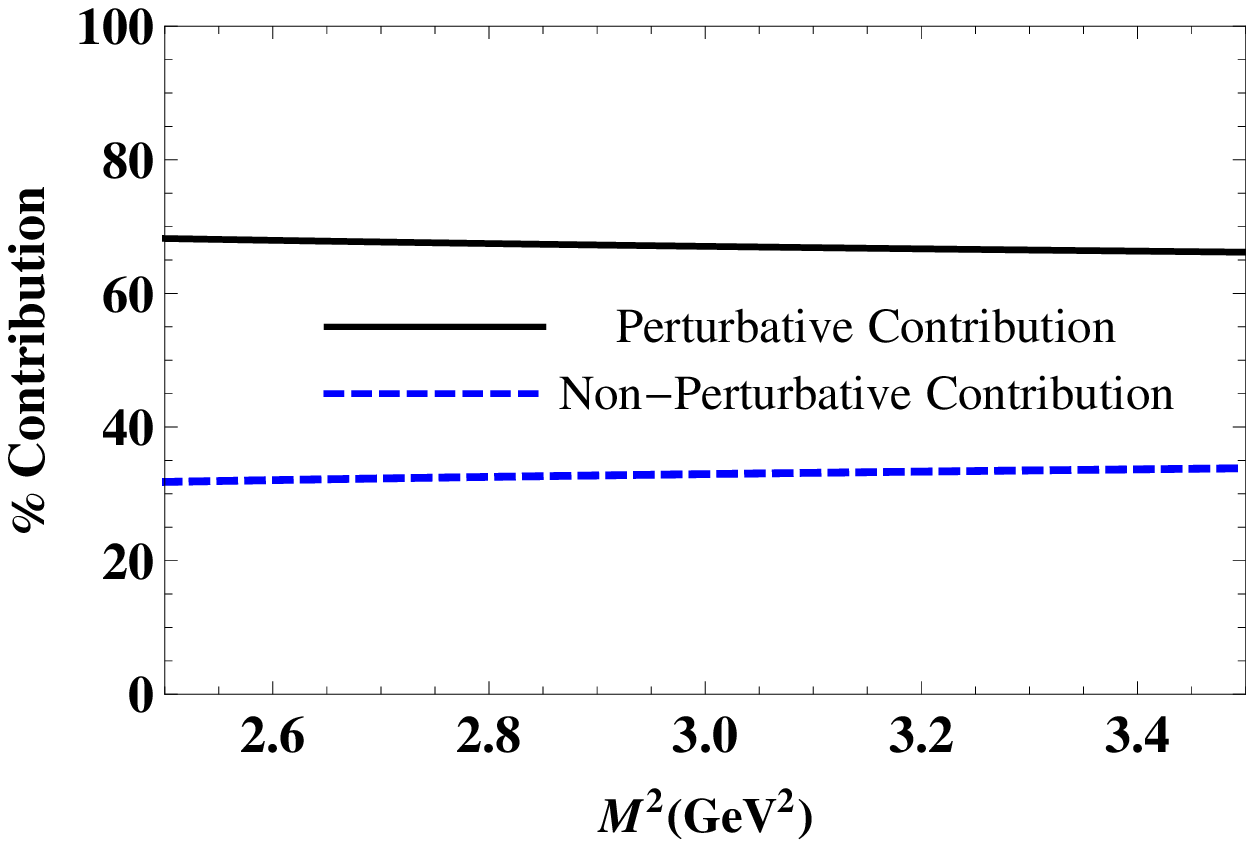}\,\, %
\includegraphics[totalheight=6cm,width=8cm]{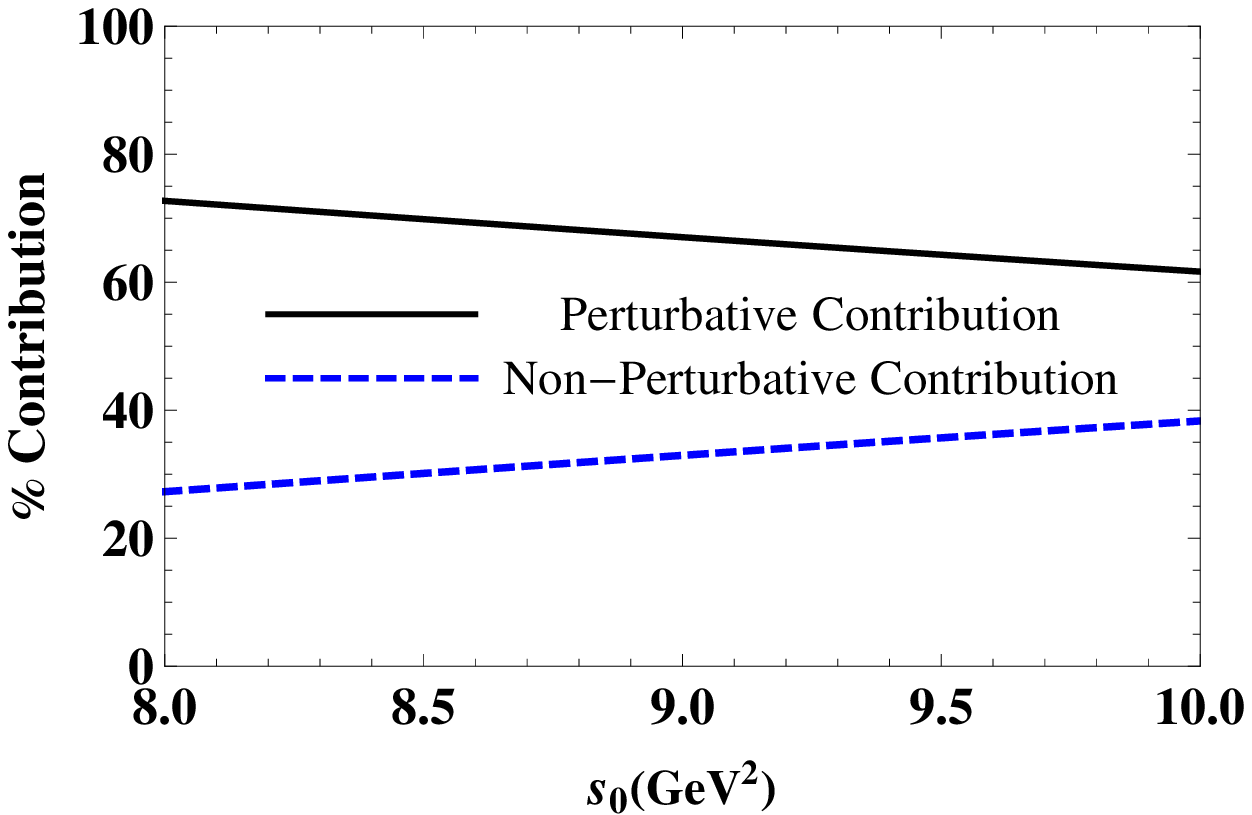}
\end{center}
\caption{ The perturbative and nonperturbative contributions to $\Pi ^{\mathrm{QCD}}(M^{2},\ s_{0})$ of the scalar particle. Left: as functions of $M^2$ at central value $s_0=9\ \mathrm{GeV}^2$, right: as functions of $s_0$ at $M^2=3\ \mathrm{GeV}^2$.}
\label{fig:Pert.NpertS}
\end{figure}

\end{widetext}From obtained sum rules for the mass and current coupling of $%
Z_{S}$ state we find
\begin{equation}
m_{Z_{S}}=2628_{-153}^{+166}\ \ \mathrm{MeV},\
f_{Z_{S}}=(0.21_{-0.05}^{+0.06})\cdot 10^{-2}\ \mathrm{GeV}^{4}.
\end{equation}%
In Figs.\ \ref{fig:Mass1} and \ref{fig:Coup1}, $m_{Z_{S}}$ and $\ f_{Z_{S}}$
are depicted as functions of $M^{2}$ and $s_{0}$. It is seen that while
effects of varying of these parameters on the mass $m_{Z_{S}}$ are small,
dependence of the current coupling $f_{Z_{S}}$ on chosen values of the
continuum threshold parameter is noticeable. These effects together with
uncertainties of the input parameters generate the theoretical errors in the
sum rule calculations, which are their unavoidable feature and may reach $%
30\%$ of the central values.
\begin{widetext}

\begin{figure}[h!]
\begin{center}
\includegraphics[totalheight=6cm,width=8cm]{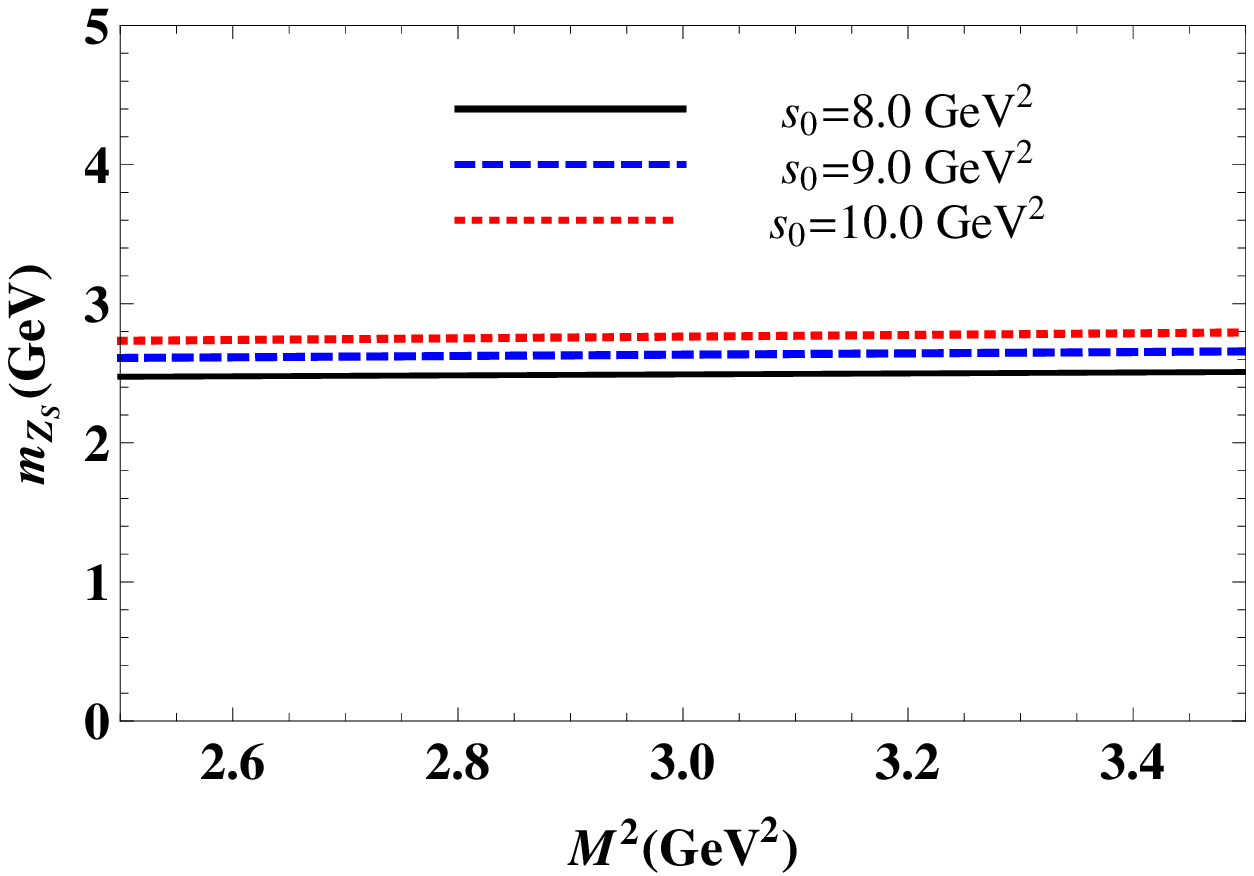}\,\, %
\includegraphics[totalheight=6cm,width=8cm]{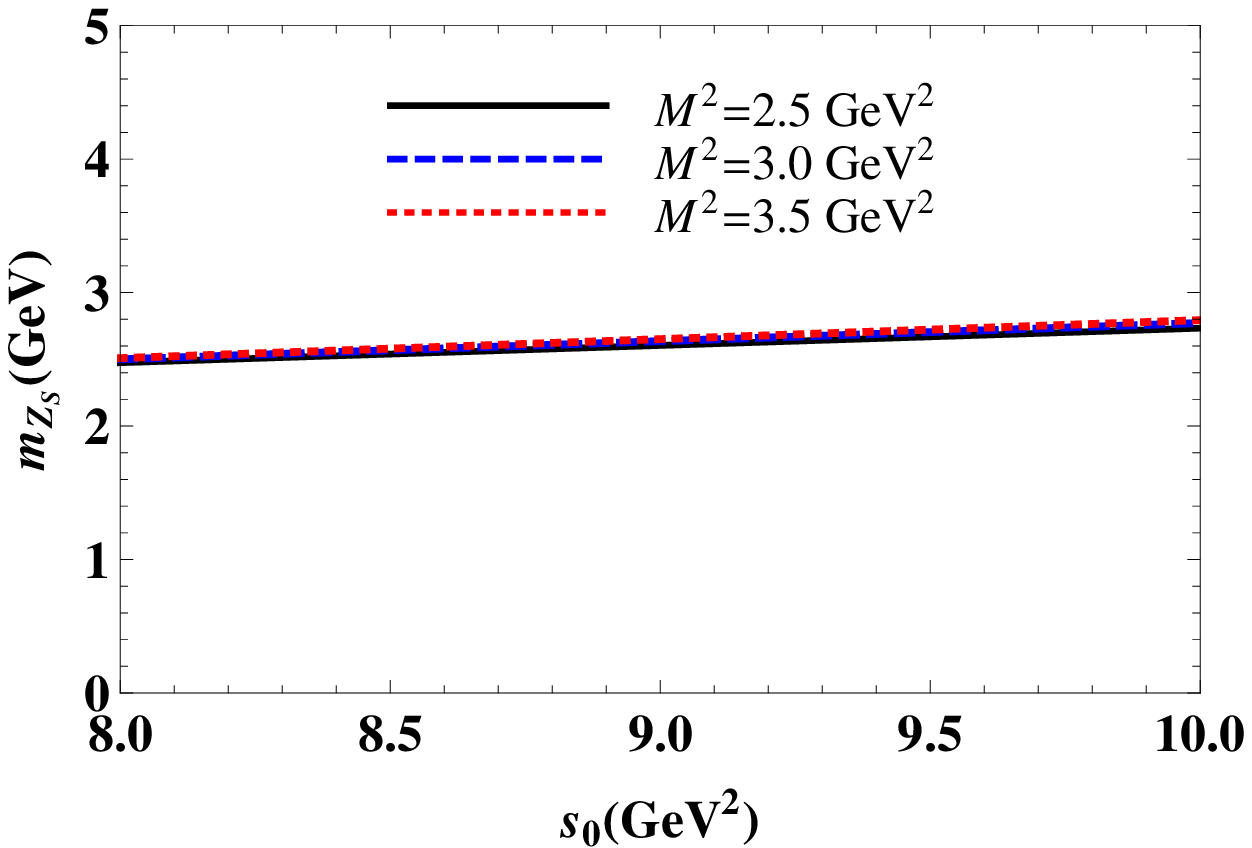}
\end{center}
\caption{ The mass of the $Z_S$ state as a function of the Borel parameter $M^2$ at fixed values of $s_0$ (left panel), and as a function of the continuum threshold $s_0$ at fixed $M^2$ (right panel).}
\label{fig:Mass1}
\end{figure}

\begin{figure}[h!]
\begin{center}
\includegraphics[totalheight=6cm,width=8cm]{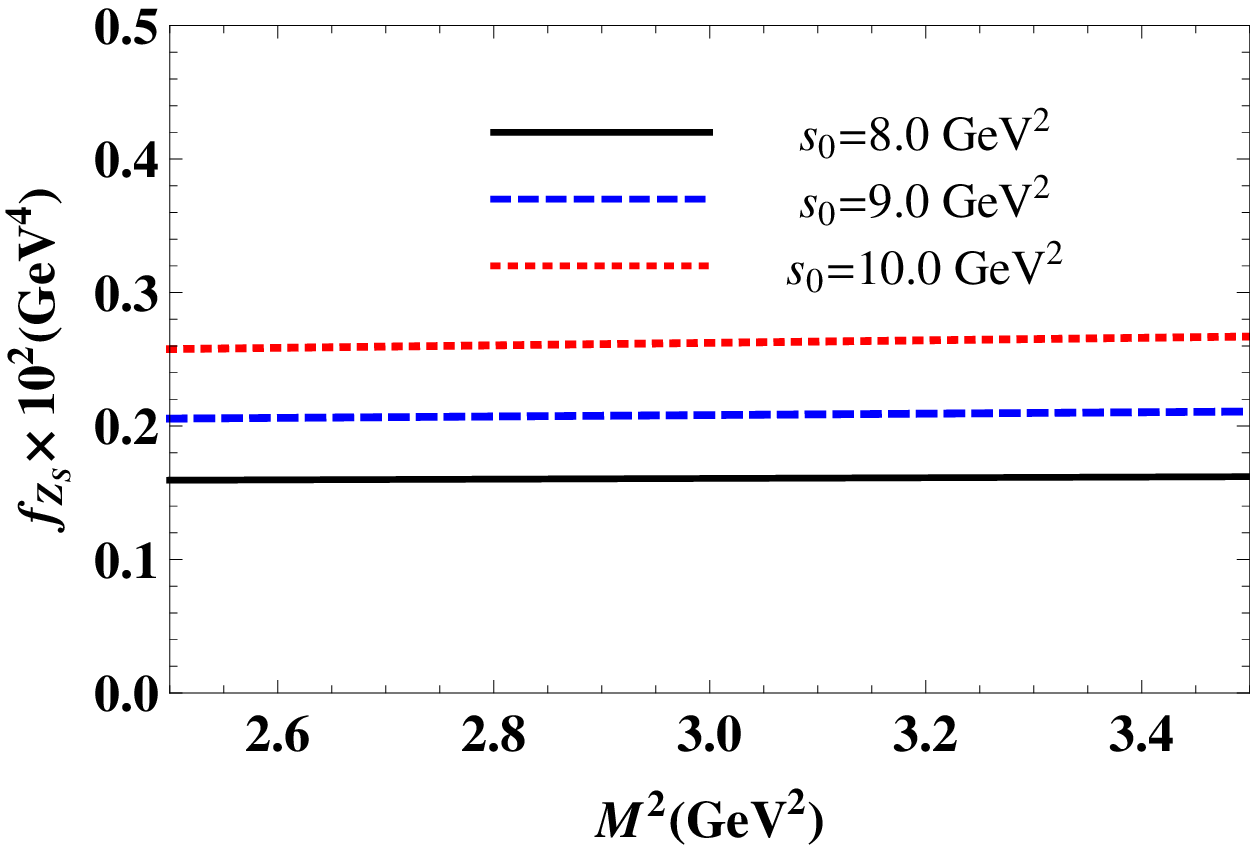}\,\, %
\includegraphics[totalheight=6cm,width=8cm]{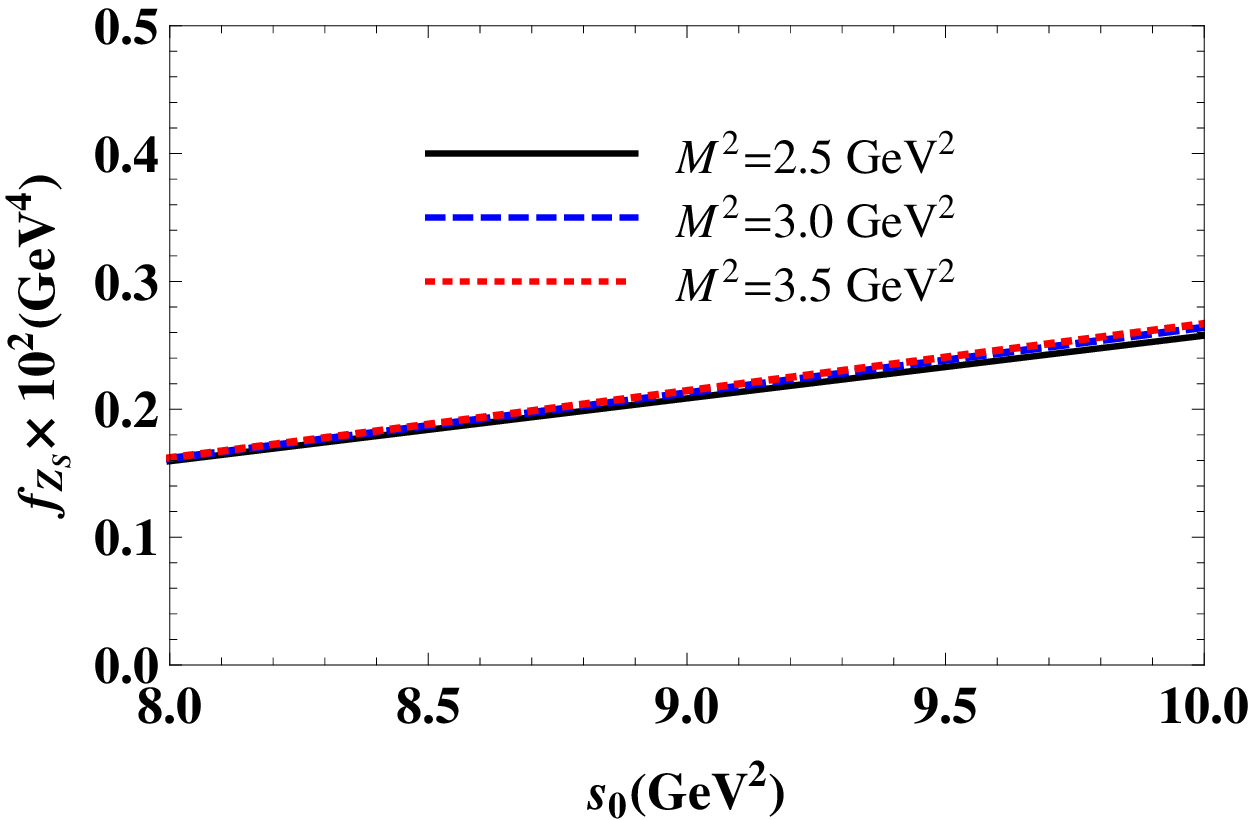}
\end{center}
\caption{ The dependence of the current coupling $f_{Z_S}$ of the
scalar $Z_S$ tetraquark on the Borel parameter at chosen values of
$s_0$ (left panel), and on the continuum threshold parameter $s_0$ at fixed $M^2$  (right
panel).} \label{fig:Coup1}
\end{figure}

\end{widetext}

\begin{figure}[h!]
\begin{center}
\includegraphics[totalheight=6cm,width=8cm]{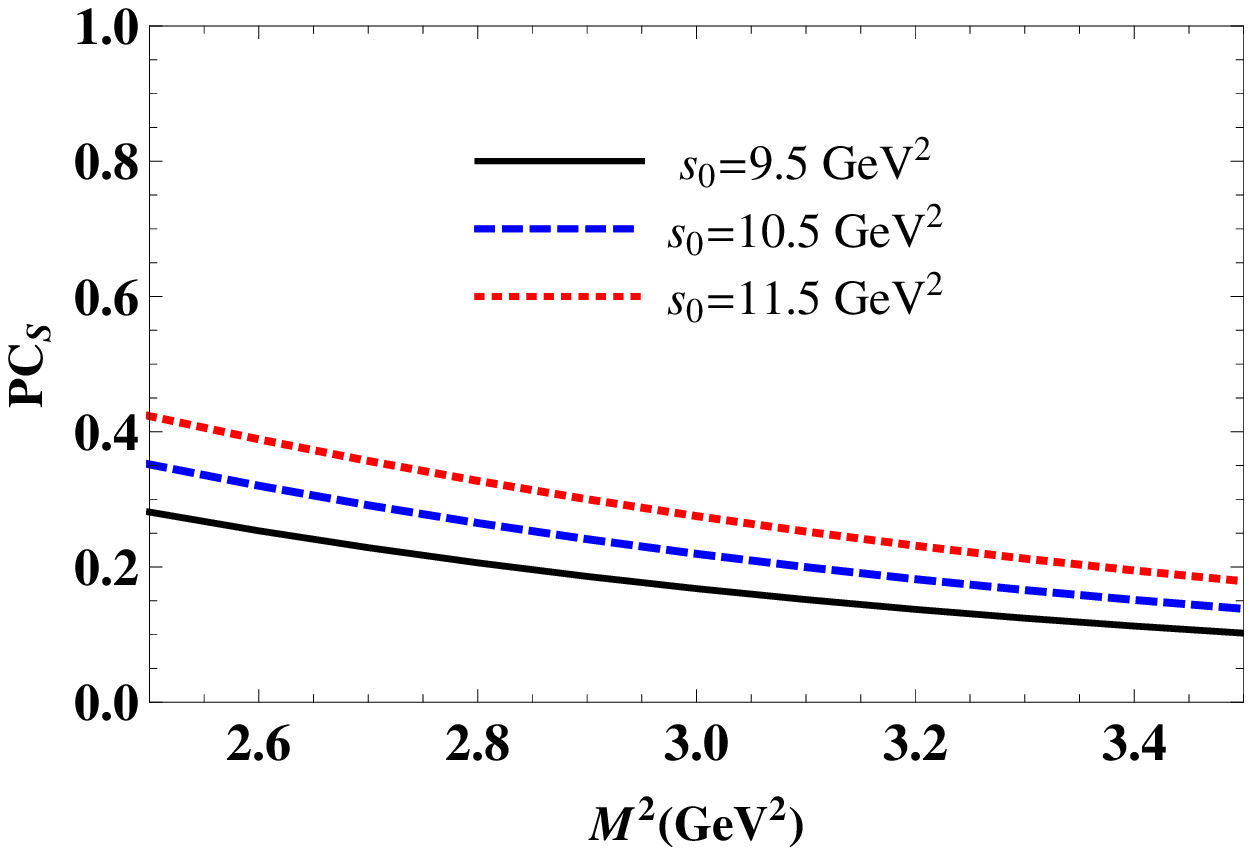}\,\,
\end{center}
\caption{ The pole contribution in the case of the axial-vector tetraquark
as a function of the Borel parameter $M^2$.}
\label{fig:PCAV}
\end{figure}

The analogous studies can be carried out for the pseudoscalar and
axial-vector tetraquarks. From performed analysis we conclude that regions
\begin{equation}
M^{2}\in \lbrack 2.5-3.5]~\mathrm{GeV}^{2},\ s_{0}~\in \lbrack 9.5-11.5]\
\mathrm{GeV}^{2}  \label{eq:BTAVparam}
\end{equation}%
can be used to evaluate the spectroscopic parameters of the pseudoscalar and
axial-vector tetraquarks, as well. Results of computations for the
axial-vector state are depicted in Figs.\ \ref{fig:PCAV}, \ref{fig:ConvAV}
and \ref{fig:Pert.NpertAV}, which confirm our conclusions. The similar
results are also valid for the pseudoscalar tetraquark $Z_{PS}$.
\begin{widetext}

\begin{figure}[h!]
\begin{center}
\includegraphics[totalheight=6cm,width=8cm]{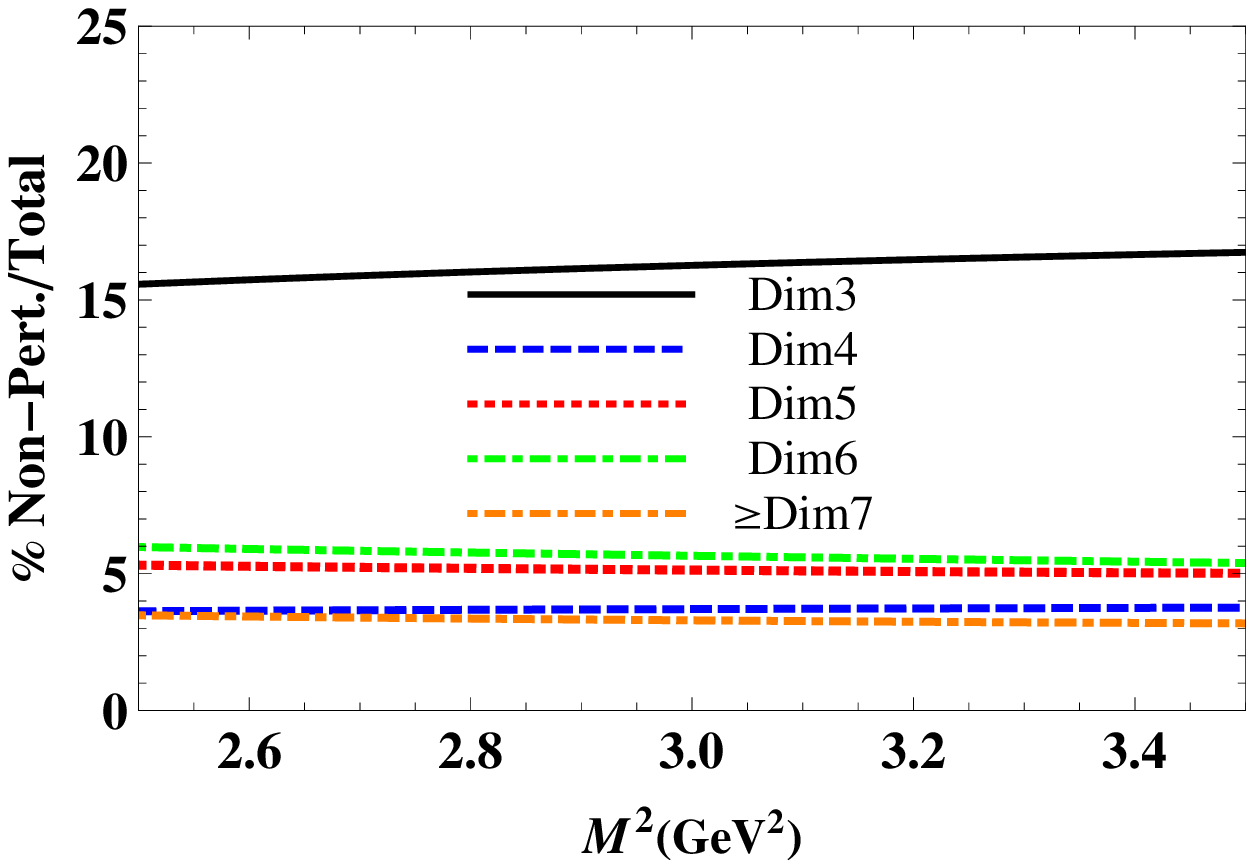}\,\, %
\includegraphics[totalheight=6cm,width=8cm]{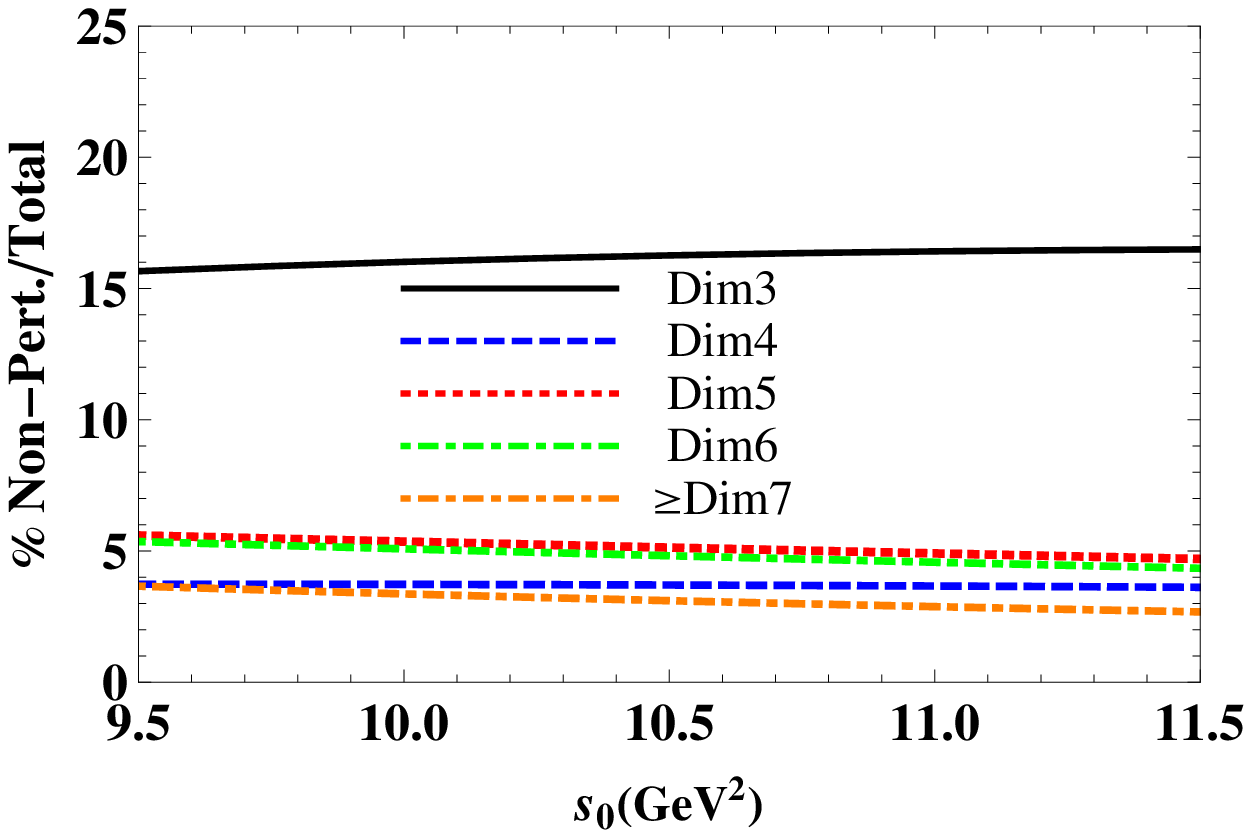}
\end{center}
\caption{ Nonperturbative contributions  to $\Pi ^{\mathrm{QCD}}(M^{2},\ s_{0})_{AV}$ as functions of $M^2$ (left panel), and $s_0$ (right
panel). } \label{fig:ConvAV}
\end{figure}

\begin{figure}[h!]
\begin{center}
\includegraphics[totalheight=6cm,width=8cm]{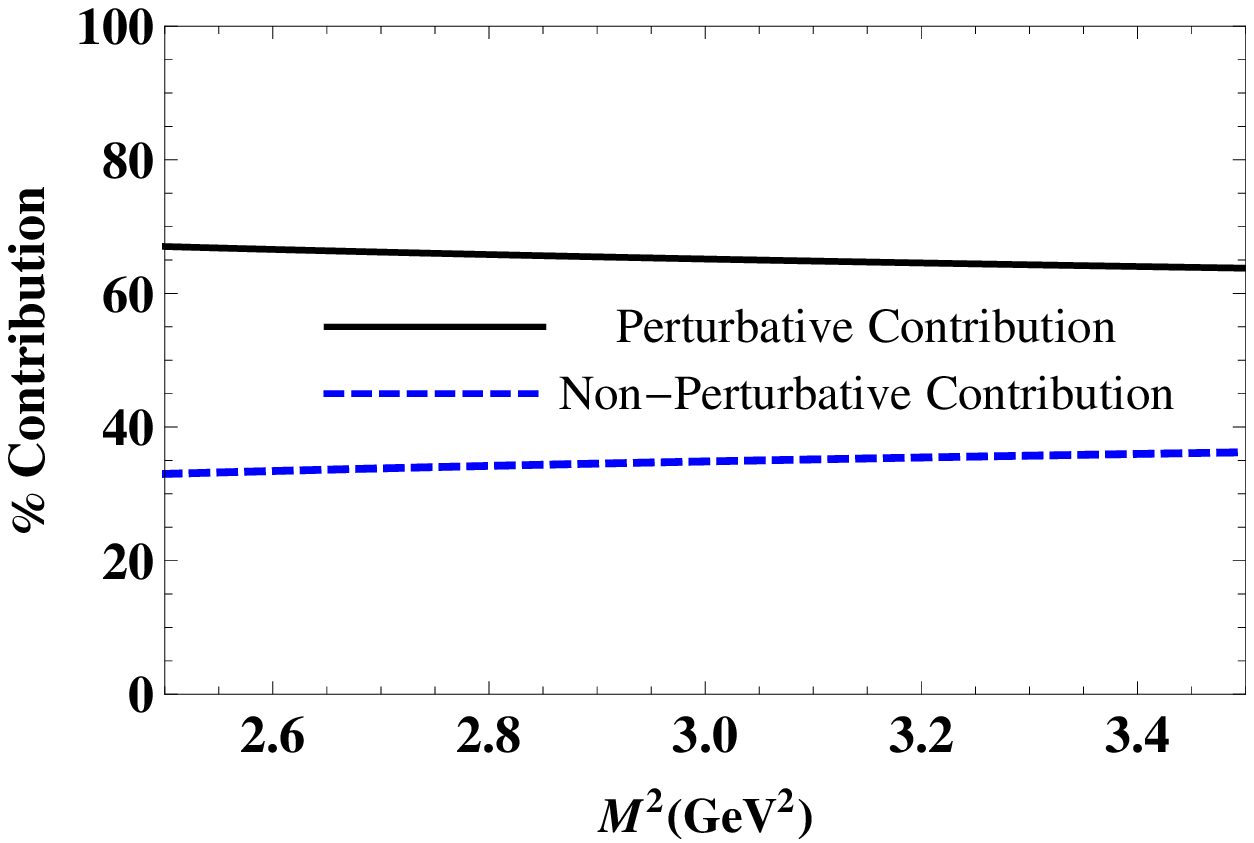}\,\, %
\includegraphics[totalheight=6cm,width=8cm]{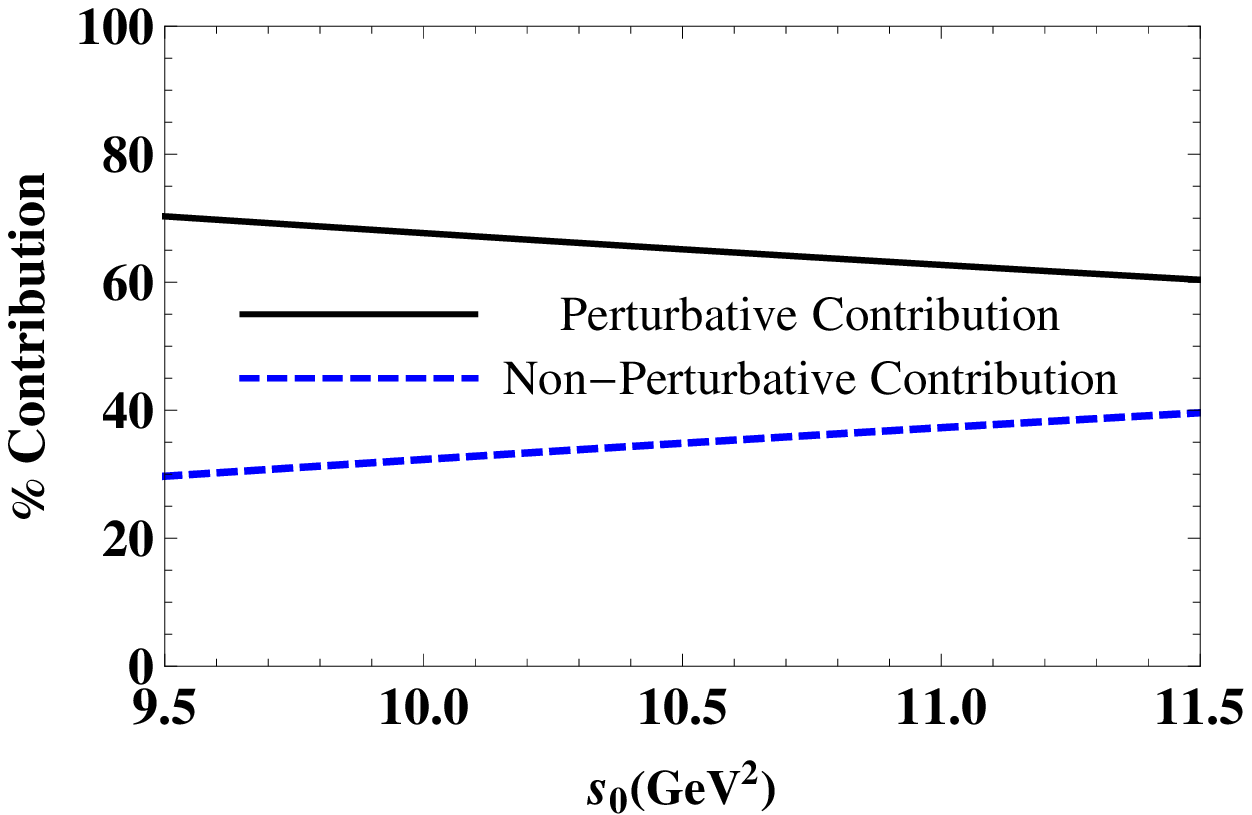}
\end{center}
\caption{ The perturbative and nonperturbative components  of $\Pi ^{\mathrm{QCD}}(M^{2},\ s_{0})_{AV}$. Left: as functions of $M^2$ at $s_0=10.5\ \mathrm{GeV}^2$, right: as functions of $s_0$ at the central value of the Borel parameter $M^2=3\ \mathrm{GeV}^2$.}
\label{fig:Pert.NpertAV}
\end{figure}

\end{widetext}

In Figs.\ \ref{fig:Mass2} and \ref{fig:Coup2} we plot dependence of the
axial-vector tetraquark's mass and current coupling on $M^{2}$ and $s_{0}$.
As is seen, estimations made for theoretical errors in the case of $Z_{S}$
are valid for the $Z_{AV}$ state, as well.

Our results for the masses and current couplings of $J^{P}=0^{+},\ 0^{-}$
and $J^{P}=1^{+}$ charm-strange tetraquarks are collected in Table \ref%
{tab:Results1}. The working ranges for the parameters $M^{2}$ and $s_{0}$,
and errors of the calculations are also presented in Table \ref{tab:Results1}%
.

\begin{widetext}

\begin{table}[tbp]
\begin{tabular}{|c|c|c|c|}
\hline\hline $Z$ & $Z_S$ & $Z_{PS}$ & $Z_{AV}$ \\ \hline\hline
$M^2 ~(\mathrm{GeV}^2$) & $2.5-3.5$ & $2.5-3.5$ & $2.5-3.5$ \\
\hline
$s_0 ~(\mathrm{GeV}^2$) & $8-10$ & $9.5-11.5$ & $9.5-11.5$ \\
\hline
$m_{Z} ~(\mathrm{MeV})$ & $2628^{+166}_{-153}$ & $2719^{+144}_{-156}$ & $%
2826^{+134}_{-157}$ \\ \hline
$f_{Z}\cdot10^{3}$ & $2.1^{+0.6}_{-0.5} ~(\mathrm{GeV}^4)$ & $%
0.83^{+0.09}_{-0.11} ~(\mathrm{GeV}^3)$ & $2.6^{+0.6}_{-0.7} ~(\mathrm{GeV}%
^4)$ \\ \hline\hline
\end{tabular}%
\caption{The masses and current couplings of the $Z_S$, $Z_{PS}$
and $Z_{AV}$ tetraquarks.} \label{tab:Results1}
\end{table}

\end{widetext}

\begin{widetext}

\begin{figure}[h!]
\begin{center}
\includegraphics[totalheight=6cm,width=8cm]{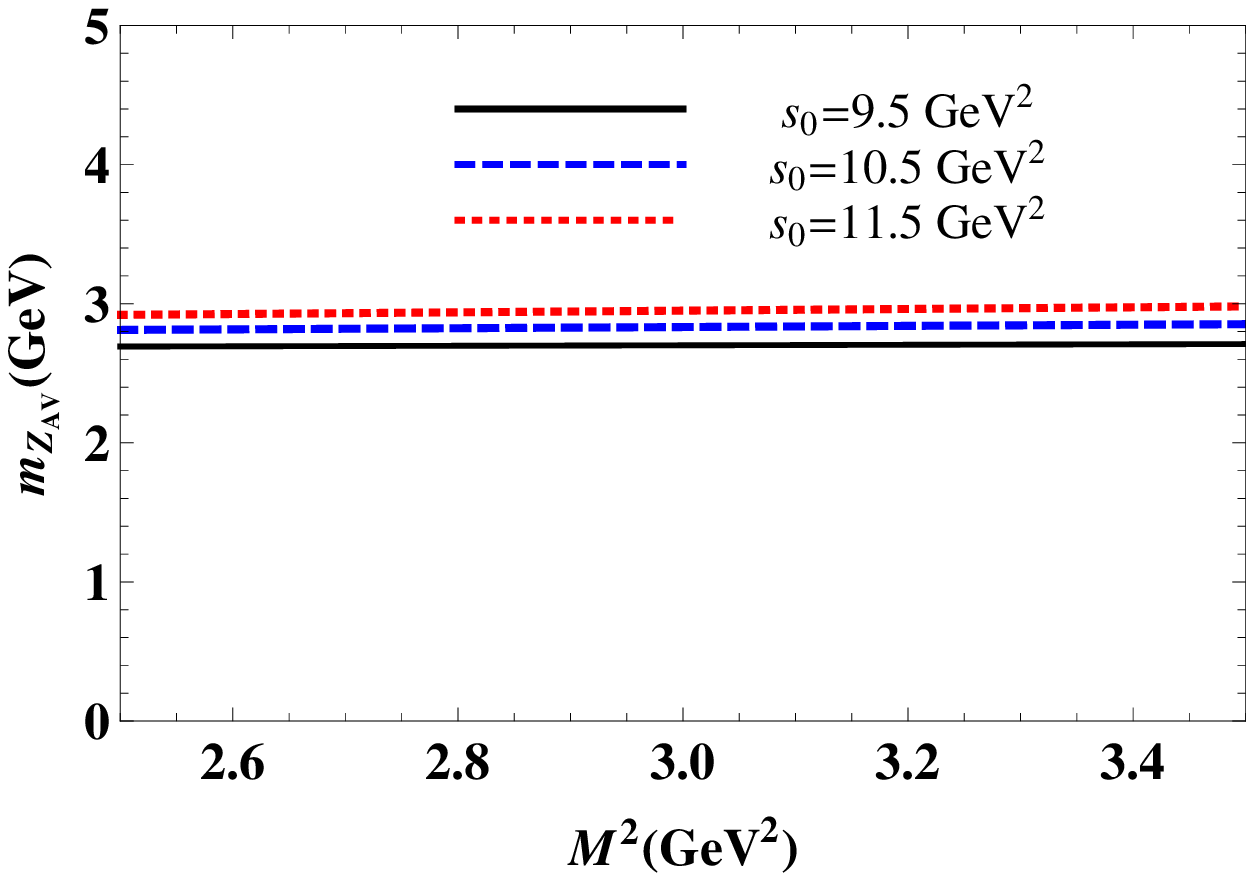}\,\, %
\includegraphics[totalheight=6cm,width=8cm]{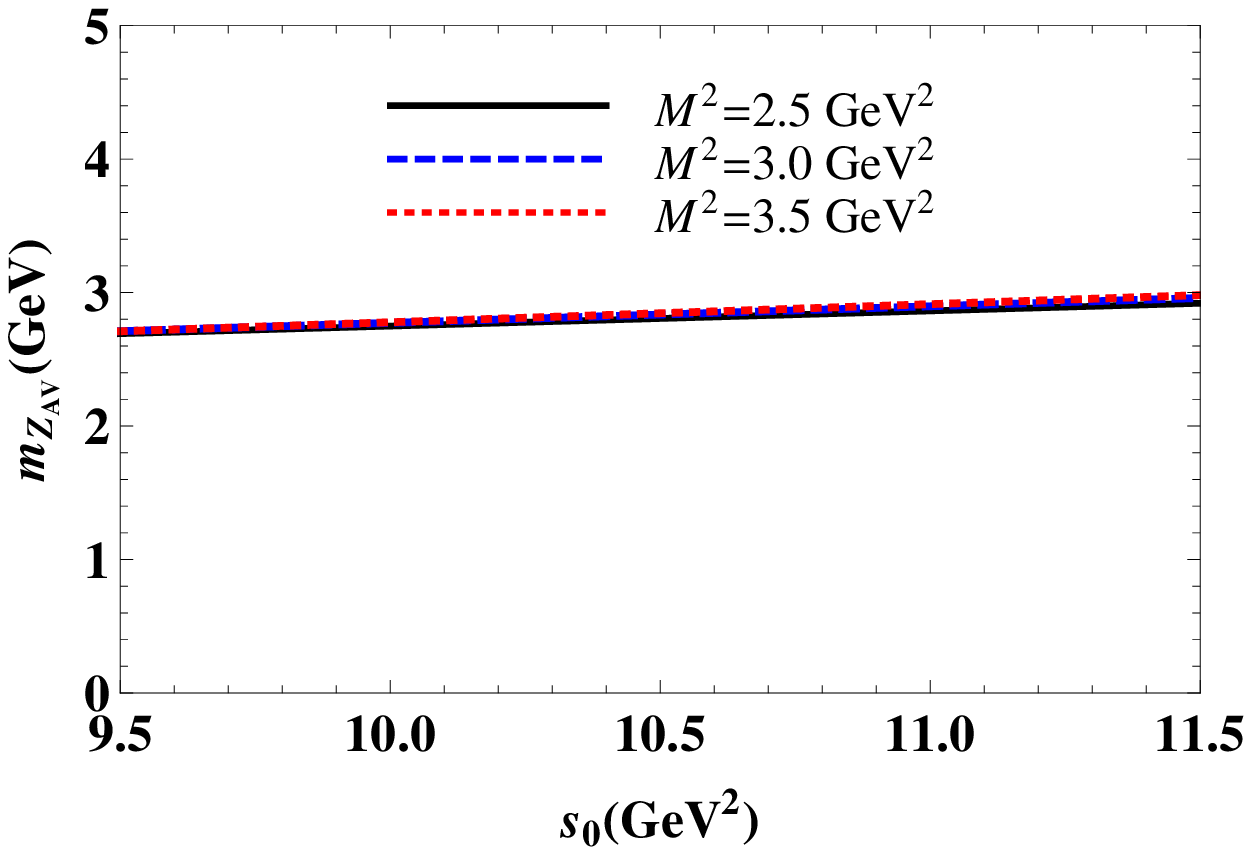}
\end{center}
\caption{ The mass of the $Z_{AV}$ state vs  Borel parameter $M^2$
at fixed values of $s_0$ (left panel), and vs continuum threshold $s_0$ at fixed values of $M^2$ (right panel).}
\label{fig:Mass2}
\end{figure}
\begin{figure}[h!]
\begin{center}
\includegraphics[totalheight=6cm,width=8cm]{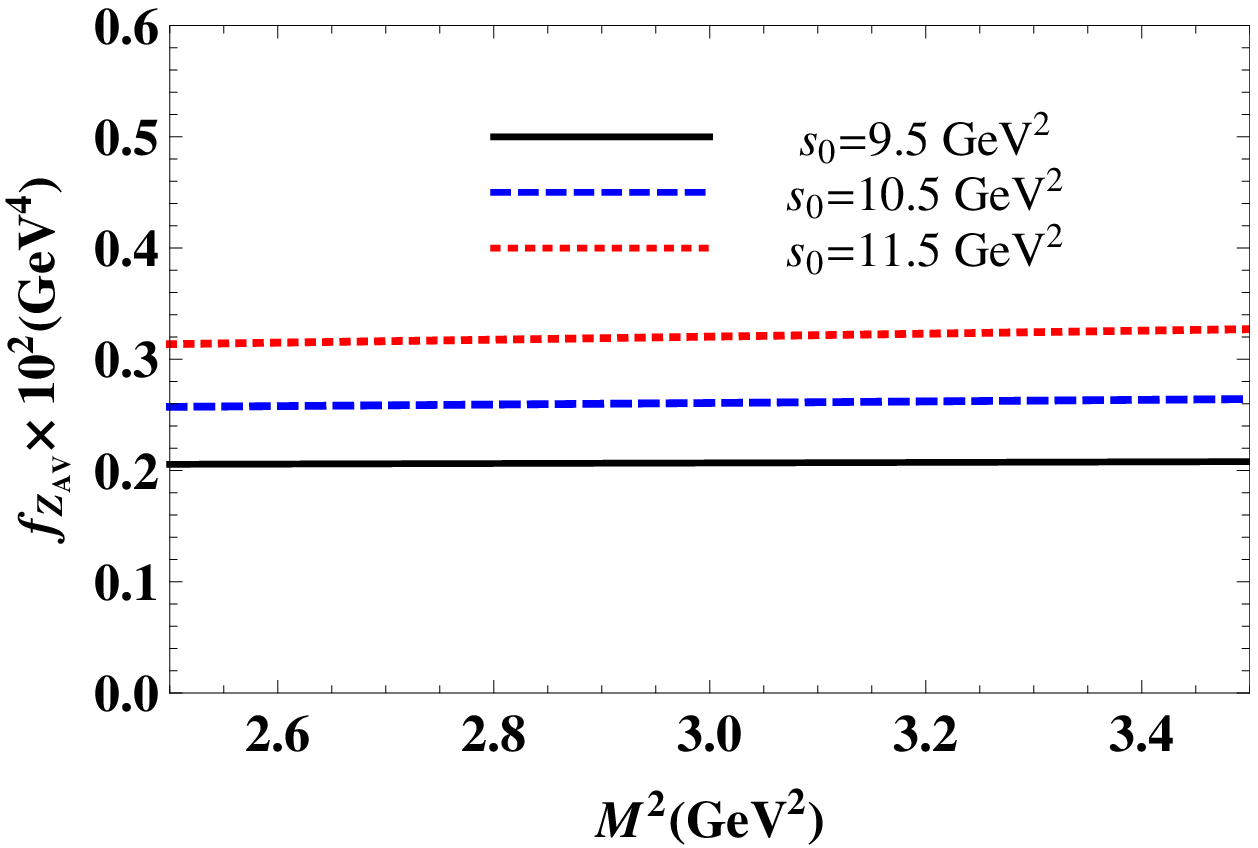}\,\, %
\includegraphics[totalheight=6cm,width=8cm]{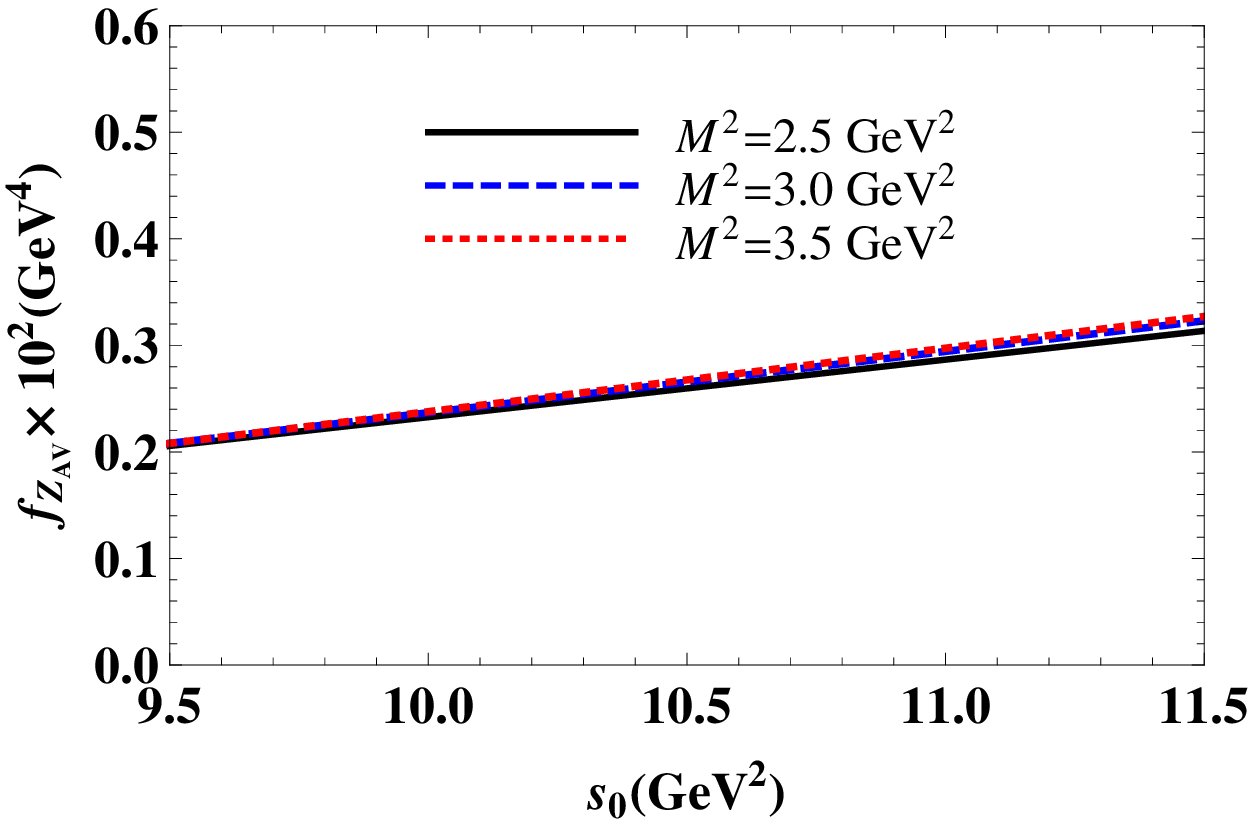}
\end{center}
\caption{ The current coupling $f_{Z_{AV}}$ of the $Z_{AV}$ state
vs Borel parameter $M^2$ at chosen values of $s_0$ (left panel),
and vs $s_0$ at fixed values of $M^2$  (right panel).} \label{fig:Coup2}
\end{figure}

\end{widetext}


\section{Decay channels of the scalar tetraquark $Z_{S}$}

\label{sec:ScalDec}
In this section we calculate the width of processes $Z_{S}\rightarrow
D_{s}\pi $, $Z_{S}\rightarrow DK$ and $Z_{S}\rightarrow D_{s1}(2460)\pi $,
which in the light of the result obtained for $m_{Z_{S}}$ are kinematically
allowed decay channels of the scalar tetraquark. It is evident that in all
these channels the final mesons are particles with negative charges. For
simplicity of expressions throughout this paper we do not show explicitly
charges of the final mesons. Let us note that first two processes are $S$%
-wave decay modes, whereas the last one is $P$-wave decay.

In order to evaluate the width of these decays we have to calculate the
strong couplings corresponding to the vertices $Z_{S}D_{s}\pi $, $Z_{S}DK$
and $Z_{S}D_{s1}(2460)\pi $. This task can be fulfilled by analysis of
corresponding correlation functions and calculating them using light-cone
sum rule method. To calculate the strong coupling $g_{Z_{S}D_{s}\pi }$ and
width of the decay $Z_{S}\rightarrow D_{s}\pi $ we consider the correlator
\begin{equation}
\Pi (p,q)=i\int d^{4}xe^{ipx}\langle \pi (q)|\mathcal{T}\{J^{D_{s}}(x)J^{%
\dag }(0)\}|0\rangle ,  \label{eq:CFScalar}
\end{equation}%
where
\begin{equation*}
J^{D_{s}}(x)=\overline{c}_{i}(x)\gamma _{5}s_{i}(x),
\end{equation*}%
is the interpolating current of the pseudoscalar meson $D_{s}$.

The correlation function $\Pi (p,q)$ in terms of the physical parameters of
the involved particles is equal to
\begin{eqnarray}
&&\Pi ^{\mathrm{Phys}}(p,q)=\frac{\langle 0|J^{D_{s}}|D_{s}\left( p\right)
\rangle }{p^{2}-m_{D_{s}}^{2}}\langle D_{s}\left( p\right) \pi
(q)|Z_{S}(p^{\prime })\rangle  \notag \\
&&\times \frac{\langle Z_{S}(p^{\prime })|J^{\dagger }|0\rangle }{p^{\prime
2}-m_{Z_{S}}^{2}}\ldots .  \label{eq:PhysScalar}
\end{eqnarray}%
By introducing the matrix elements
\begin{eqnarray}
&&\langle 0|J^{D_{s}}|D_{s}\left( p\right) \rangle =\frac{%
m_{D_{s}}^{2}f_{D_{s}}}{m_{c}+m_{s}},  \notag \\
&&\langle D_{s}\left( p\right) \pi (q)|Z_{S}(p^{\prime })\rangle
=g_{Z_{S}D_{s}\pi }p\cdot p^{\prime },  \label{eq:Vertex1}
\end{eqnarray}%
we can rewrite $\Pi ^{\mathrm{Phys}}(p,q)$ in the form%
\begin{equation*}
\Pi ^{\mathrm{Phys}}(p,q)=\frac{g_{Z_{S}D_{s}\pi
}m_{D_{s}}^{2}m_{Z_{S}}f_{D_{s}}f_{Z_{S}}}{(p^{2}-m_{D_{s}}^{2})(p^{\prime
2}-m_{Z_{S}}^{2})(m_{c}+m_{s})}p\cdot p^{\prime }+\ldots
\end{equation*}%
Here $m_{D_{s}}$ and $f_{D_{s}}$ are the mass and decay constant of the
meson $D_{s}$, respectively.

In terms of the quark-gluon degrees of freedom $\Pi (p,q)$ is given by the
expression
\begin{eqnarray}
&&\Pi ^{\mathrm{QCD}}(p,q)=i\int d^{4}xe^{ipx}\left\{ \left[ \gamma _{5}%
\widetilde{S}_{s}^{ia}(x)\gamma _{5}\widetilde{S}_{c}^{bi}(-x)\gamma _{5}%
\right] _{\alpha \beta }\right.  \notag \\
&&\times \langle \pi (q)|\overline{d}_{\alpha }^{b}(0)u_{\beta
}^{a}(0)|0\rangle -\left[ \gamma _{5}\widetilde{S}_{s}^{ia}(x)\gamma _{5}%
\widetilde{S}_{c}^{ai}(-x)\gamma _{5}\right]  \notag \\
&&\left. \times \langle \pi (q)|\overline{d}_{\alpha }^{b}(0)u_{\beta
}^{b}(0)|0\rangle \right\} ,  \label{eq:QCD3}
\end{eqnarray}%
where $\alpha $ and $\beta $ are the spinor indices. To continue we employ
the expansion
\begin{equation}
\overline{u}_{\alpha }^{a}d_{\beta }^{d}\rightarrow \frac{1}{4}\Gamma
_{\beta \alpha }^{j}\left( \overline{u}^{a}\Gamma ^{j}d^{d}\right) ,
\label{eq:MatEx}
\end{equation}%
with $\Gamma ^{j}$ being the full set of Dirac matrices
\begin{equation*}
\Gamma ^{j}=\mathbf{1,\ }\gamma _{5},\ \gamma _{\lambda },\ i\gamma
_{5}\gamma _{\lambda },\ \sigma _{\lambda \rho }/\sqrt{2}.
\end{equation*}%
The operators $\overline{u}^{a}(0)\Gamma ^{j}d^{d}(0)$, as well as
three-particle operators that appear due to insertion of $G_{\mu \nu }$ from
propagators $\widetilde{S}_{s}(x)$ and $\widetilde{S}_{c}(-x){}$ into $%
\overline{u}^{a}(0)\Gamma ^{j}d^{d}(0)$ give rise to local matrix elements
of the pion. In other words, instead of the distribution amplitudes the
function $\Pi ^{\mathrm{QCD}}(p,q)$ depends on the pion's local matrix
elements. Then, the conservation of four-momentum in the
tetraquark-meson-meson vertex can be obeyed by setting $q=0$. In the limit $%
q\rightarrow 0$ we get $p=p^{\prime }$ and have to carry out Borel
transformations over one variable $p^{2}$. This condition has to be
implemented in the physical side of the sum rule, as well \ \cite%
{Agaev:2016dev,Agaev:2017foq}.

After substituting Eq.\ (\ref{eq:MatEx}) into the expression of the
correlation function and performing the summation over color indices in
accordance with recipes presented in a detailed form in Ref.\ \cite%
{Agaev:2016dev}, we fix local matrix elements that enter to $\Pi ^{\mathrm{%
QCD}}(p,q)$. It turns out that only the matrix element of the pion
\begin{equation}
\langle 0|\overline{d}(0)i\gamma _{5}u(0)|\pi (q)\rangle =f_{\pi }\mu _{\pi
},  \label{eq:MatE2}
\end{equation}%
where $\mu _{\pi }=-2\langle \bar{q}q\rangle /f_{\pi }^{2}$ contributes to
the correlation function. Other matrix elements including three-particle
ones either do not contribute to a final expression of $\Pi ^{\mathrm{QCD}%
}(p,q)$ or vanish in the soft limit $q\rightarrow 0$.

In the soft limit the Borel transformation of relevant invariant function $%
\Pi ^{\mathrm{QCD}}(p^{2})$ can be obtained after the following operations:
we find the spectral density $\rho ^{\mathrm{pert.}}(s)$ as imaginary part
of $\Pi ^{\mathrm{pert.}}(p^{2})$, which is the perturbative component of
the full correlation function. It is calculated using Eq.\ (\ref{eq:QCD3})
and keeping in the quark propagators only their perturbative components. All
other terms in $\Pi ^{\mathrm{QCD}}(p^{2})$ constitute the nonperturbative
peace of the correlator, i.e. function $\Pi ^{\mathrm{n.-pert.}}(p^{2})$. We
calculate Borel transformation of $\Pi ^{\mathrm{n.-pert.}}(p^{2})$ directly
from Eq.\ (\ref{eq:QCD3}) in accordance with prescriptions of Ref.\ \cite%
{Belyaev:1994zk} , and by this way bypass intermediate steps, i. e.
computation of $\rho ^{\mathrm{n.-pert.}}(s)$, which becomes unnecessary in
this case. This approach considerably simplifies calculations and allows us
to find explicitly $\Pi ^{\mathrm{n.-pert.}}(M^{2})\equiv \mathcal{B}\Pi ^{%
\mathrm{n.-pert.}}(p^{2})$. For the spectral density $\rho ^{\mathrm{pert.}%
}(s)$ we obtain%
\begin{equation}
\rho ^{\mathrm{pert.}}(s)=\frac{f_{\pi }\mu _{\pi }}{16\pi ^{2}s}%
(s-m_{c}^{2})(s+2m_{c}m_{s}-m_{c}^{2}).  \label{eq:SD1}
\end{equation}%
The Borel transformed $\Pi ^{\mathrm{n.-pert.}}(M^{2})$ contains terms up to
nine dimensions and reads%
\begin{equation}
\Pi ^{\mathrm{n.-pert.}}(M^{2})=\frac{f_{\pi }\mu _{\pi }}{12M^{2}}%
e^{-m_{c}^{2}/M^{2}}\sum_{l=1}^{5}F_{l}(M^{2}),  \label{eq:Borel1}
\end{equation}%
where%
\begin{eqnarray}
&&F_{1}(M^{2})=-\langle \bar{s}s\rangle
(m_{c}^{2}m_{s}+2m_{c}M^{2}-m_{s}M^{2}),  \notag \\
&&F_{2}(M^{2})=\frac{m_{c}}{12M^{2}}\langle \frac{\alpha _{s}G^{2}}{\pi }%
\rangle \int_{0}^{1}\frac{dze^{m_{c}^{2}/M^{2}-m_{c}^{2}/[M^{2}z(1-z)]}}{%
z(z-1)^{3}}  \notag \\
&&\times \left[ m_{c}^{3}z+2m_{s}M^{2}(1-z)^{2}z+m_{c}^{2}m_{s}(1-z)\right] ,
\notag \\
&&F_{3}(M^{2})=\frac{m_{c}^{3}}{6M^{4}}\langle \overline{s}g_{s}\sigma
Gs\rangle \left( m_{c}m_{s}+3M^{2}\right) ,  \notag \\
&&F_{4}(M^{2})=-\frac{m_{c}\pi ^{2}}{18M^{6}}\langle \frac{\alpha _{s}G^{2}}{%
\pi }\rangle \langle \bar{s}s\rangle \left[ m_{c}^{3}m_{s}\right.  \notag \\
&&\left. +2m_{c}M^{2}(m_{c}+m_{s})+6M^{4}\right] ,  \notag \\
&&F_{5}(M^{2})=\frac{m_{c}\pi ^{2}}{108M^{10}}\langle \frac{\alpha _{s}G^{2}%
}{\pi }\rangle \langle \overline{s}g_{s}\sigma Gs\rangle \left(
m_{c}m_{s}+3M^{2}\right)  \notag \\
&&\times \left( m_{c}^{4}+6m_{c}^{2}M^{2}+6M^{4}\right) .
\label{eq:Ffunctions}
\end{eqnarray}

Then in the soft limit the Borel transformation of $\Pi ^{\mathrm{QCD}%
}(p^{2})$ takes the form%
\begin{eqnarray}
&&\mathcal{B}\Pi ^{\mathrm{QCD}}(p^{2})=\int_{(m_{c}+m_{s})^{2}}^{\infty
}\rho ^{\mathrm{pert.}}(s)e^{-s/M^{2}}ds  \notag \\
&&+\Pi ^{\mathrm{n.-pert.}}(M^{2}).
\end{eqnarray}%
In the same limit the Borel transformation of $\Pi ^{\mathrm{Phys}}(p^{2})$
is given by the expression%
\begin{equation}
\mathcal{B}\Pi ^{\mathrm{Phys}}(p^{2})=\frac{g_{Z_{S}D_{s}\pi
}m_{D_{s}}^{2}m_{Z_{S}}f_{D_{s}}f_{Z_{S}}}{m_{c}+m_{s}}m^{2}\frac{%
e^{-m^{2}/M^{2}}}{M^{2}}+...,  \label{eq:PhysScalA}
\end{equation}%
where $m^{2}=(m_{D_{s}}^{2}+m_{Z_{S}}^{2})/2.$

After equating $\mathcal{B}\Pi ^{\mathrm{Phys}}(p^{2})$ and $\mathcal{B}\Pi
^{\mathrm{QCD}}(p^{2})$ one has to subtract contributions of higher
resonances and continuum states. In the case of standard sum rules (i.e. $%
q\neq 0$) this can be carried out quite easily, because Borel transformation
suppress all undesired terms in the physical side of the equality. But in
the soft limit $\mathcal{B}\Pi ^{\mathrm{Phys}}(p^{2})$ contains terms which
are not suppressed even after Borel transformation \cite{Belyaev:1994zk},
therefore additional manipulations are required to remove them from the
phenomenological side of the sum rule. Acting by the operator
\begin{equation}
\mathcal{P}(M^{2},m^{2})=\left( 1-M^{2}\frac{d}{dM^{2}}\right)
M^{2}e^{m^{2}/M^{2}}  \label{eq:softop}
\end{equation}%
one can achieve this goal \cite{Ioffe:1983ju}. Then subtraction can be
performed in a standard manner and leads to the sum rule%
\begin{eqnarray}
&&g_{Z_{S}D_{s}\pi }=\frac{m_{c}+m_{s}}{%
m_{D_{s}}^{2}m_{Z_{S}}m^{2}f_{D_{s}}f_{Z_{S}}}\mathcal{P}(M^{2},m^{2})
\notag \\
&&\times \left[ \int_{(m_{c}+m_{s})^{2}}^{s_{0}}\rho ^{\mathrm{pert.}%
}(s)e^{-s/M^{2}}ds+\Pi ^{\mathrm{n.-pert.}}(M^{2})\right] .  \notag \\
&&{}  \label{eq:SRcoupl1}
\end{eqnarray}%
It is worth noting that we do not perform continuum subtractionin in
nonperturbative terms $\sim (M^{2})^{0}$ and $\sim (M^{2})^{-n},\
n=1,2\ldots $ \cite{Belyaev:1994zk}.

With the coupling $g_{Z_{S}D_{s}\pi }$ at hands it is straightforward to
evaluate the width of the decay $Z_{S}\rightarrow D_{s}\pi $%
\begin{eqnarray}
&&\Gamma (Z_{S}\to D_{s}\pi )=\frac{g_{Z_{S}D_{s}\pi }^{2}}{96\pi
m_{Z_{S}}^{2}}\left( m_{Z_{S}}^{2}+m_{D_{s}}^{2}-m_{\pi }^{2}\right) ^{2}
\notag \\
&&\times f(m_{Z_{S}},m_{D_{s}},m_{\pi }),
\end{eqnarray}%
where the function $f(x,y,z)$ is%
\begin{equation}
f(x,y,z)=\frac{1}{2x}\sqrt{%
x^{4}+y^{4}+z^{4}-2x^{2}y^{2}-2x^{2}z^{2}-2y^{2}z^{2}}.
\end{equation}

Another decay channel of the doubly charged charm-strange tetraquark is $%
Z_{S}\rightarrow DK$. Correlation function that should be considered in this
case is given by the expression
\begin{equation}
\Pi _{K}(p,q)=i\int d^{4}xe^{ipx}\langle K(q)|\mathcal{T}\{J^{D}(x)J^{\dag
}(0)\}|0\rangle ,
\end{equation}%
where the interpolating current for $D$ meson is
\begin{equation*}
J^{D}(x)=\overline{c}_{i}(x)\gamma _{5}d_{i}(x).
\end{equation*}%
The analysis of the channel $Z_{S}\rightarrow DK$ does not differ
considerably from consideration of $Z_{S}\rightarrow D_{s}\pi $ decay.
Because particles in final states $D_{s}\pi $ and $DK$ are pseudoscalar
mesons, differences between two decay channels are encoded in the matrix
element
\begin{equation*}
\langle 0|J^{D}|D\left( p\right) \rangle =\frac{m_{D}^{2}f_{D}}{m_{c}+m_{d}},
\end{equation*}%
and local matrix element of $K$ meson
\begin{equation}
\langle 0|\overline{u}(0)i\gamma _{5}s(0)|K(q)\rangle =\frac{f_{K}m_{K}^{2}}{%
m_{s}+m_{u}},  \label{eq:MatEl3}
\end{equation}%
that contributes to $\Pi _{K}^{\mathrm{QCD}}(p,q)$, where $m_{K}$ and $f_{K}$
are the mass and decay constant of $K$ meson. The strong coupling $%
g_{Z_{S}DK}$ with evident replacements is defined by Eq.\ (\ref{eq:Vertex1}).

In the $P$-wave decay $Z_{S}\rightarrow D_{s1}(2460)\pi $ the interpolating
current, matrix element and strong coupling of the axial-vector meson $%
D_{s1}(2460)$ are introduced by means of the formulas
\begin{eqnarray}
&&J_{\mu }^{D_{s1}}(x)=\overline{c}_{i}(x)\gamma _{\mu }\gamma _{5}s_{i}(x),
\notag \\
&&\langle 0|J_{\mu }^{D_{s1}}|D_{s1}\left( p\right) \rangle
=f_{D_{s1}}m_{D_{s1}}\varepsilon _{\mu },  \notag \\
&&\langle D_{s1}\left( p\right) \pi (q)|Z_{S}(p^{\prime })\rangle
=g_{Z_{S}D_{s1}\pi }\varepsilon ^{\ast }\cdot p^{\prime },
\label{eq:Vertex2}
\end{eqnarray}%
with $m_{D_{s1}}$ , $f_{D_{s1}}$ and $\varepsilon _{\mu }$ being its mass ,
decay constant and polarization vector, respectively.

The remaining operations and intermediate steps in both cases are standard
ones, therefore we refrain from presenting them here in a detailed form, and
write down only formula for the decay width $\Gamma (Z_{S}\rightarrow
D_{s1}(2460)\pi ):$
\begin{equation}
\Gamma (Z_{S}\rightarrow D_{s1}(2460)\pi )=\frac{g_{Z_{S}D_{s1}\pi }^{2}}{%
24\pi m_{D_{s1}}^{2}}f^{3}(m_{Z_{S}},m_{D_{s1}},m_{\pi }).
\end{equation}

Numerical calculations are carried out using the sum rules derived for
strong couplings and expressions for widths of different decay modes of $%
Z_{S}$. The masses and decay constants of $D_{s}$ , $D$ and $D_{s1}(2460)$,
as well as $\pi $ and $K$ mesons which we employ in numerical computations
are collected in Table\ \ref{tab:Param}. The masses of particles are taken
from Ref.\ \cite{Olive:2016xmw}, for decay constants of $D$ and $D_{s}$
mesons we use information from Ref.\ \cite{Rosner:2015wva}, decay constant
of $D_{s1}(2460)$ is borrowed from \cite{Sungu:2010zz}. Table\ \ref%
{tab:Param} contains also parameters of the $D^{\ast }$, $D_{s}^{\ast }$ and
$D_{s0}^{\ast }(2317)$ mesons which will be used in the next section.

The Borel parameter $M^{2}$ and continuum threshold $s_{0}$ in coupling
calculations are chosen as in Eq. \ (\ref{eq:BTparam}). For the strong
couplings of the explored vertices and width of the decay modes we obtain:
for the channel $Z_{S}\rightarrow D_{s}\pi $
\begin{eqnarray}
&&g_{Z_{S}D_{s}\pi } =(0.51\pm 0.14)\ \mathrm{GeV}^{-1}  \notag \\
&&\Gamma (Z_{S} \rightarrow D_{s}\pi )=(8.27\pm 2.32)\ \mathrm{MeV,}
\label{eq:DW1}
\end{eqnarray}%
for the mode $Z_{S}\rightarrow DK$%
\begin{eqnarray}
&&g_{Z_{S}DK} =(1.54\pm 0.43)\ \mathrm{GeV}^{-1},  \notag \\
&&\Gamma (Z_{S} \rightarrow DK)=(57.41\pm 14.93)\ \mathrm{MeV,}
\label{eq:DW2}
\end{eqnarray}%
and for $Z_{S}\rightarrow D_{s1}(2460)\pi $%
\begin{eqnarray}
&&g_{Z_{S}Ds1\pi } =26.16\pm 7.36,  \notag \\
&&\Gamma (Z_{S} \rightarrow D_{s1}(2460)\pi )=(1.21\pm 0.38)\ \mathrm{MeV.}
\label{eq:DW3}
\end{eqnarray}%
The full width of the scalar tetraquark $Z_{S}$ on the basis of considered
decay modes is equal to%
\begin{equation*}
\Gamma _{Z_{S}}=(66.89\pm 15.11)\ \ \mathrm{MeV,}
\end{equation*}%
which is typical for a diquark-antidiquark state: The tetraquark $Z_{S}$
belongs neither to a class of broad resonances $\Gamma \sim 200\ \mathrm{MeV}
$ nor to a class of very narrow states $\Gamma \sim 1\ \mathrm{MeV}$.

The charmed particle composed of four different quarks as a partner of the $%
X(5568)$ resonance was previously investigated in our work \cite%
{Agaev:2016lkl}. We analyzed this state using the interpolating currents of
both $C\gamma _{5}\otimes \gamma _{5}C$ and $C\gamma _{\mu }\otimes \gamma
^{\mu }C$ types. The diquark-antidiquark composition of $X_{c}=[su][%
\overline{c}\overline{d}]$ means that it is a neutral particle.
Nevertheless, it is instructive to compare parameters of $X_{c}$ with
results for $Z_{S}$ obtained in the present work. In the case of the
interpolating current $C\gamma _{5}\otimes \gamma _{5}C$ we found $%
m_{X_{c}}=(2634\pm 62)\ \mathrm{MeV}$ which is very close to our present
result. The processes $X_{c}\rightarrow \overline{D}^{0}\overline{K}^{0}$
and $X_{c}\rightarrow D_{s}^{-}\pi ^{+}$ were also subject of studies in
Ref. \cite{Agaev:2016lkl}. Width of these decay channels $\Gamma
(X_{c}\rightarrow \overline{D}^{0}\overline{K}^{0})=(53.7\pm 11.6)\ \mathrm{%
MeV}$ and $\Gamma (X_{c}\rightarrow D_{s}^{-}\pi ^{+})=(8.2\pm 2.1)\ \mathrm{%
MeV}$ are comparable with ones presented in Eqs.\ (\ref{eq:DW1}) and (\ref%
{eq:DW2}).
\begin{table}[tbp]
\begin{tabular}{|c|c|}
\hline\hline
Parameters & Values \\ \hline\hline
$m_{D}$ & $(1869.5 \pm 0.4) ~\mathrm{MeV}$ \\
$f_{D}$ & $(211.9 \pm 1.1)~\mathrm{MeV}$ \\
$m_{D_s}$ & $(1969.0 \pm 1.4) ~\mathrm{MeV}$ \\
$f_{D_s}$ & $(249.0 \pm 1.2) ~\mathrm{MeV}$ \\
$m_{D_{s1}} $ & $(2459.6 \pm 0.9)~\mathrm{MeV} $ \\
$f_{D_{s1}}$ & $(481\pm 164) ~\mathrm{MeV}$ \\
$m_{D_{s}^{\ast }}$ & $(2112.1\pm 0.4)~\mathrm{MeV}$ \\
$f_{D_{s}^{\ast }}$ & $(308\pm 21) ~\mathrm{MeV}$ \\
$m_{D^{\ast}}$ & $(2010.26\pm 0.25)~\mathrm{MeV}$ \\
$f_{D^{\ast }}$ & $(252.2\pm 22.66) ~\mathrm{MeV}$ \\
$m_{D_{s0}^{\ast }}$ & $(2318.0\pm 1.0)~\mathrm{MeV}$ \\
$f_{D_{s0}^{\ast }}$ & $201\ \mathrm{MeV}$ \\
$m_{K}$ & $(493.677\pm 0.016) ~\mathrm{MeV}$ \\
$f_{K}$ & $156~\mathrm{MeV}$ \\
$m_{\pi}$ & $(139.57061 \pm 0.00024 )~\mathrm{MeV}$ \\
$f_{\pi}$ & $131~\mathrm{MeV}$ \\ \hline\hline
\end{tabular}%
\caption{Parameters of the mesons used in numerical calculations.}
\label{tab:Param}
\end{table}


\section{ $Z_{PS}\rightarrow \ D_{s}^{\ast }\protect\pi ,\ D^{\ast }K,\
D_{s0}^{\ast}(2317)\protect\pi $ and $Z_{AV}\rightarrow \ D_{s}^{\ast }%
\protect\pi ,\ D^{\ast }K,\ D_{s1}(2460)\protect\pi $ decays of the
pseudoscalar and axial-vector tetraquarks}

\label{sec:PsVDec}
The pseudoscalar $Z_{PS}$ and axial-vector $Z_{AV}$ tetraquarks may decay
through different channels. Among kinematically allowed decay channels of $%
Z_{PS}$ state are $S-$wave mode $Z_{PS}\rightarrow D_{s0}(2317)\pi $, and $%
P- $wave modes $Z_{PS}\rightarrow D_{s}^{\ast }\pi$ and $\ D^{\ast }K$. The
decays of the tetraquark $Z_{AV}$ include $S-$wave channels $%
Z_{AV}\rightarrow D_{s}^{\ast }\pi ,\ D^{\ast }K$ and $P-$wave mode $%
Z_{AV}\rightarrow D_{s1}(2460)\pi$.

It is seen that both $Z_{PS}$ and $Z_{AV}$ states decay to $D_{s}^{\ast }\pi
$ and $D^{\ast }K$, therefore these channels should be analyzed in a
connected form. We start our investigation from analysis of the decays $%
Z_{AV}\rightarrow D_{s}^{\ast }\pi $ and $Z_{PS}\rightarrow D_{s}^{\ast }\pi
$, and construct the following correlation function
\begin{equation}
\Pi _{\mu \nu }(p,q)=i\int d^{4}xe^{ipx}\langle \pi (q)|\mathcal{T}\{J_{\mu
}^{D_{s}^{\ast }}(x)J_{\nu }^{\dag }(0)\}|0\rangle ,  \label{eq:CF4}
\end{equation}%
where $J_{\mu }^{D_{s}^{\ast }}(x)$ is the interpolating current of the $%
D_{s}^{\ast }$ meson
\begin{equation}
J_{\mu }^{D_{s}^{\ast }}(x)=\overline{c}_{l}(x)\gamma _{\mu }s_{l}(x).
\label{eq:Curr3}
\end{equation}%
The function $\Pi _{\mu \nu }(p,q)$ will be computed employing QCD sum rule
on the light-cone and using a technique of the soft-meson approximation. \
Because the current $J_{\nu }(x)$ couples to both the pseudoscalar and
axial-vector tetraquarks the correlator $\Pi _{\mu \nu }^{\mathrm{Phys}%
}(p,q) $ expressed in terms of the physical parameters of the involved
particles and vertices contains two components: Indeed, for $\Pi _{\mu \nu
}^{\mathrm{Phys}}(p,q)$ we find:
\begin{eqnarray}
&&\Pi _{\mu \nu }^{\mathrm{Phys}}(p,q)=\frac{\langle 0|J_{\mu }^{D_{s}^{\ast
}}|D_{s}^{\ast }\left( p\right) \rangle }{p^{2}-m_{D_{s}^{\ast }}^{2}}%
\langle D_{s}^{\ast }\left( p\right) \pi (q)|Z_{PS}(p^{\prime })\rangle
\notag \\
&&\times \frac{\langle Z_{PS}(p^{\prime })|J_{\nu }^{\dagger }|0\rangle }{%
p^{\prime 2}-m_{Z_{PS}}^{2}}+\frac{\langle 0|J_{\mu }^{D_{s}^{\ast
}}|D_{s}^{\ast }\left( p\right) \rangle }{p^{2}-m_{D_{s}^{\ast }}^{2}}%
\langle D_{s}^{\ast }\left( p\right) \pi (q)|Z_{AV}(p^{\prime })\rangle
\notag \\
&&\times \frac{\langle Z_{AV}(p^{\prime })|J_{\nu }^{\dagger }|0\rangle }{%
p^{\prime 2}-m_{Z_{AV}}^{2}}\ldots .  \label{eq:CF5}
\end{eqnarray}%
The terms in Eq. (\ref{eq:CF5}) are contributions of vertices $%
Z_{PS}D_{s}^{\ast }\pi $ and $Z_{AV}D_{s}^{\ast }\pi $, where all particles
are on their ground states. The dots stand for effects due to the higher
resonances and continuum.

We introduce the $D_{s}^{\ast }$ meson matrix element
\begin{equation*}
\langle 0|J_{\mu }^{D_{s}^{\ast }}|D_{s}^{\ast }\left( p\right) \rangle
=f_{D_{s}^{\ast }}m_{D_{s}^{\ast }}\varepsilon _{\mu },
\end{equation*}%
where $m_{D_{s}^{\ast }}$ , $f_{D_{s}^{\ast }}$ and $\varepsilon _{\mu }$
are its mass, decay constant and polarization vector, respectively. We
define also the matrix elements corresponding to the vertices in the
following manner
\begin{eqnarray}
&&\langle D_{s}^{\ast }\left( p\right) \pi (q)|Z_{AV}(p^{\prime })\rangle
=g_{Z_{AV}D_{s}^{\ast }\pi }\left[ \left( p\cdot p^{\prime }\right) \left(
\varepsilon ^{\ast }\cdot \varepsilon ^{\prime }\right) \right.  \notag \\
&&\left. -\left( q\cdot \varepsilon ^{\prime }\right) \left( p^{\prime
}\cdot \varepsilon ^{\ast }\right) \right] ,  \label{eq:ME1}
\end{eqnarray}%
and%
\begin{equation}
\langle D_{s}^{\ast }\left( p\right) \pi (q)|Z_{PS}(p^{\prime })\rangle
=g_{Z_{PS}D_{s}^{\ast }\pi }p^{\prime }\cdot \varepsilon .  \label{eq:ME2}
\end{equation}%
After some manipulations the ground state terms in $\Pi _{\mu \nu }^{\mathrm{%
Phys}}(p,q)$ can be easily rewritten as:
\begin{eqnarray}
&&\Pi _{\mu \nu }^{\mathrm{Phys}}(p,q)=g_{Z_{AV}D_{s}^{\ast }\pi }\frac{%
m_{D_{s}^{\ast }}f_{D_{s}^{\ast }}m_{Z_{AV}}f_{Z_{AV}}}{\left(
p^{2}-m_{D_{s}^{\ast }}^{2}\right) \left( p^{\prime 2}-m_{Z_{AV}}^{2}\right)
}  \notag \\
&&\times \left( \frac{m_{Z_{AV}}^{2}+m_{D_{s}^{\ast }}^{2}}{2}g_{\mu \nu
}-p_{\mu }p_{\nu }^{\prime }\right)  \notag \\
&&+\frac{g_{Z_{PS}D_{s}^{\ast }\pi }f_{D_{s}^{\ast }}m_{Z_{PS}}f_{Z_{PS}}}{%
\left( p^{2}-m_{D_{s}^{\ast }}^{2}\right) \left( p^{\prime
2}-m_{Z_{PS}}^{2}\right) m_{D_{s}^{\ast }}}  \notag \\
&&\times \frac{m_{Z_{PS}}^{2}-m_{D_{s}^{\ast }}^{2}}{2}p_{\mu }p_{\nu
}^{\prime }+\ldots .  \label{eq:CF5A}
\end{eqnarray}%
One sees that $\Pi _{\mu \nu }^{\mathrm{Phys}}(p,q)$ contains two structures
$\sim g_{\mu \nu }$ and $\sim p_{\mu }p_{\nu }^{\prime }.$ The same
structures appear in the second part of the sum rule which is the
correlation function Eq.\ (\ref{eq:CF4}) calculated in terms of quark
propagators. For $\Pi _{\mu \nu }^{\mathrm{QCD}}(p,q)$ we get
\begin{eqnarray}
&&\Pi _{\mu \nu }^{\mathrm{QCD}}(p,q)=i\int d^{4}xe^{ipx}\left\{ \left[
\gamma _{\nu }\widetilde{S}_{s}^{ia}(x){}\gamma _{\mu }\right. \right.
\notag \\
&&\left. \times \widetilde{S}_{c}^{bi}(-x){}\gamma _{5}\right] _{\alpha
\beta }\langle \pi (q)|\overline{d}_{\alpha }^{b}(0)u_{\beta
}^{a}(0)|0\rangle  \notag \\
&&\left. -\left[ \gamma _{\nu }\widetilde{S}_{s}^{ia}(x){}\gamma _{\mu }%
\widetilde{S}_{c}^{ai}(-x){}\gamma _{5}\right] \langle \pi (q)|\overline{d}%
_{\alpha }^{b}(0)u_{\beta }^{b}(0)|0\rangle \right\} .  \label{eq:CF6}
\end{eqnarray}%
We use invariant amplitudes corresponding to structures $\sim g_{\mu \nu }$
from $\Pi _{\mu \nu }^{\mathrm{Phys}}(p,q)$ and $\Pi _{\mu \nu }^{\mathrm{QCD%
}}(p,q)$ to derive sum rule for the coupling $g_{Z_{AV}D_{s}^{\ast }\pi }.$
To this end, we equate these invariant amplitudes and carry out calculations
in accordance with scheme described in rather detailed form in the previous
section. Obtained by this way sum rule is employed to evaluate the strong
coupling $g_{Z_{AV}D_{s}^{\ast }\pi }$. It is utilized as an input parameter
at the second stage of analysis, when we employ invariant amplitudes
corresponding to structures $\sim p_{\mu }p_{\nu }^{\prime }$ to derive sum
rule for $g_{Z_{PS}D_{s}^{\ast }\pi }$.

The decays $Z_{AV}\rightarrow D^{\ast }K$ and $Z_{PS}\rightarrow D^{\ast }K$
can be investigated in the same way, but one has to start from the
correlator
\begin{equation}
\Pi _{\mu \nu }(p,q)=i\int d^{4}xe^{ipx}\langle K(q)|\mathcal{T}\{J_{\mu
}^{D^{\ast }}(x)J_{\nu }^{\dag }(0)\}|0\rangle ,  \label{eq:CF7}
\end{equation}%
with $J_{\mu }^{D^{\ast }}(x)$%
\begin{equation}
J_{\mu }^{D^{\ast }}(x)=\overline{c}_{l}(x)\gamma _{\mu }d_{l}(x)
\label{eq:Curr4}
\end{equation}%
The remaining analysis does not differ from calculations of the decays $%
Z_{AV}\rightarrow D_{s}^{\ast }\pi $ and $Z_{PS}\rightarrow D_{s}^{\ast }\pi$%
, and therefore we do not provide further details.

There are also two processes $Z_{PS}\rightarrow D_{s0}^{\ast }(2317)\pi $
and $Z_{AV}\rightarrow D_{s1}(2460)\pi $ which are not connected with each
other, and can be studied separately. Let us consider, for example, decay $%
Z_{PS}\rightarrow D_{s0}^{\ast }(2317)\pi $ that can be explored by means of
the correlator
\begin{equation}
\Pi _{\nu }(p,q)=i\int d^{4}xe^{ipx}\langle \pi (q)|\mathcal{T}%
\{J^{D_{s0}^{\ast }}(x)J_{\nu }^{\dag }(0)\}|0\rangle ,  \label{eq:CF8}
\end{equation}%
where the interpolating current $J_{\mu }^{D_{s0}^{\ast }}(x)$ is chosen in
the form%
\begin{equation}
J^{D_{s0}^{\ast }}(x)=\overline{c}^{i}(x)s^{i}(x).  \label{eq:Curr5}
\end{equation}%
The correlation function $\Pi _{\nu }(p,q)$ has the following
phenomenological representation
\begin{eqnarray}
\Pi _{\nu }^{\mathrm{Phys}}(p,q) &=&\frac{\langle 0|J^{D_{s0}^{\ast
}}|D_{s0}^{\ast }\left( p\right) \rangle }{p^{2}-m_{D_{s0}^{\ast }}^{2}}%
\langle D_{s0}^{\ast }\left( p\right) \pi (q)|Z_{PS}(p^{\prime })\rangle
\notag \\
&&\times \frac{\langle Z_{PS}(p^{\prime })|J_{\nu }^{\dagger }|0\rangle }{%
p^{\prime 2}-m_{Z_{PS}}^{2}}+\ldots .
\end{eqnarray}%
Using of the matrix element%
\begin{equation}
\langle 0|J^{D_{s0}^{\ast }}|D_{s0}^{\ast }\left( p\right) \rangle
=f_{D_{s0}^{\ast }}m_{D_{s0}^{\ast }},
\end{equation}%
and also the vertex%
\begin{equation}
\langle D_{s0}^{\ast }\left( p\right) \pi (q)|Z_{PS}(p^{\prime })\rangle
=g_{Z_{PS}D_{s0}^{\ast }\pi }p\cdot p^{\prime },
\end{equation}%
it can be rewritten as
\begin{equation}
\Pi _{\nu }^{\mathrm{Phys}}(p,q)=g_{Z_{PS}D_{s0}^{\ast }\pi }\frac{%
f_{Z_{PS}}m_{Z_{PS}}f_{D_{s0}^{\ast }}m_{D_{s0}^{\ast }}}{(p^{2}-m^{2})^{2}}%
m^{2}p_{\nu }^{\prime }+\ldots ,  \label{eq:CFDS0}
\end{equation}%
where $m^{2}=(m_{Z_{PS}}^{2}+m_{D_{s0}^{\ast }}^{2})/2$.  In order to match
the obtained expression with the same structure from $\Pi _{\nu }^{\mathrm{%
QCD}}(p,q)$ we keep in Eq.\ (\ref{eq:CFDS0}) dependence on $p_{\nu }^{\prime
}$, whereas in the invariant amplitude, i. e. in the function $\sim p_{\nu
}^{\prime }$ implement the soft limit.

The same correlation function $\Pi _{\nu }(p,q)$ in terms of quark
propagators and pion's matrix elements is given by formula%
\begin{eqnarray}
&&\Pi _{\nu }^{\mathrm{QCD}}(p,q)=i\int d^{4}xe^{ipx}\left\{ \left[ \gamma
_{\nu }\widetilde{S}_{s}^{ia}(x)\widetilde{S}_{c}^{bi}(-x)\gamma _{5}\right]
_{\alpha \beta }\right.   \notag \\
&&\times \langle \pi (q)|\overline{d}_{\alpha }^{b}(0)u_{\beta
}^{a}(0)|0\rangle -\left[ \gamma _{\nu }\widetilde{S}_{s}^{ia}(x){}%
\widetilde{S}_{c}^{ai}(-x){}\gamma _{5}\right]   \notag \\
&&\left. \times \langle \pi (q)|\overline{d}_{\alpha }^{b}(0)u_{\beta
}^{b}(0)|0\rangle \right\} .
\end{eqnarray}%
After calculations one finds that in $\Pi _{\nu }^{\mathrm{QCD}}(p,q)$
survives only the structure $\sim p_{\nu }^{\prime }$. By equating invariant
amplitudes from both sides and performing all manipulations it is possible
to derive the sum rule for the coupling $g_{Z_{PS}D_{s0}^{\ast }\pi }$. The
similar analysis has been carried out for the decay $Z_{AV}\rightarrow
D_{s1}(2460)\pi $, as well.

In numerical calculations of the $Z_{PS}$ and $Z_{AV}$ states' strong
couplings the Borel parameter and continuum threshold are chosen within the
same ranges as in computations of their masses (see, Table\ \ref%
{tab:Results1}). As input parameters we employ also mass and decay constant
of the mesons $D_{s}^{\ast },\ D^{\ast }$ and $\ D_{s0}^{\ast }(2317)$ from
Table\ \ref{tab:Param}. It is worth noting that the decay constants $%
f_{D_{s}^{\ast }}$, $f_{D^{\ast }}$ and $f_{D_{s0}^{\ast }}$ have been taken
from Refs.\ \cite{Agaev:2015faa,Lucha:2014xla,Narison:2003td}, respectively.

Results for strong couplings and width of decay modes of $Z_{PS}$ and $%
Z_{AV} $ tetraquarks are presented in Table\ \ref{tab:Results2}. Using these
predictions one can evaluate full widths of the pseudoscalar and
axial-vector tetraquarks $Z_{PS}$ and $Z_{AV}$:
\begin{equation}
\Gamma_{Z_{PS}}=(38.1\pm 7.1)~\mathrm{MeV},
\end{equation}
and
\begin{equation}
\Gamma_{Z_{AV}}=(47.3\pm 11.1)~\mathrm{MeV}.
\end{equation}
As is seen, the tetraquarks $Z_{PS}$ and $Z_{AV}$ are narrower than the
scalar state $Z_S$. Nevertheless, we cannot classify them as narrow
resonances.

\begin{widetext}

\begin{table}[tbp]
\begin{tabular}{|c|c|c|}
\hline\hline
Decay & Strong couplings & Decay Width  \\
\hline\hline
$Z_{AV} \to D_{s}^{\ast }\pi $ & $(0.26 \pm 0.07)~\mathrm{GeV}^{-1}$ & $(7.94 \pm 2.21)~\mathrm{MeV}$ \\
$Z_{AV} \to D^{\ast } K$ & $(0.63 \pm 0.17) ~\mathrm{GeV}^{-1}$ & $(37.38 \pm 10.84)~\mathrm{MeV}$ \\
$Z_{AV}\to D_{s1}\pi$ & $(1.55 \pm 0.43) ~\mathrm{GeV}^{-1}$ &  $(2.02 \pm 0.59)~\mathrm{MeV}$ \\
$Z_{PS}\to D_{s}^{\ast }\pi $ & $3.18 \pm 0.94$ &$(4.37 \pm 1.27)~\mathrm{MeV}$ \\
$Z_{PS}\to D^{\ast }K$ & $8.24 \pm 2.39 $ &$(19.09 \pm 5.73)~\mathrm{MeV}$ \\
$Z_{PS} \to D_{s0}^{\ast}\pi$ & $(0.76 \pm 0.18) ~\mathrm{GeV}^{-1}$ & $(14.64 \pm 3.94)~\mathrm{MeV}$\\
\hline\hline
\end{tabular}%
\caption{The strong couplings and decay widths of the $Z_{AV}$ and
$Z_{PS}$ tetraquarks. } \label{tab:Results2}
\end{table}

\end{widetext}


\section{Conclusions}

\label{sec:Concl}
In the present work we have investigated the charm-strange tetraquarks $Z_{%
\overline{c}s}=[sd][\overline{u}\overline{c}]$ by calculating their
spectroscopic parameters and decay channels. It is easy to see that these
states bear two units of electric charge $-|e|$ and belong to a class of
doubly charged tetraquarks. Their counterparts with the structure $Z_{c%
\overline{s}}=[uc][\overline{s}\overline{d}]$ have evidently a charge $+2|e|$%
. We have considered scalar, pseudoscalar and axial-vector doubly charged
states. Their masses have been obtained using QCD two-point sum rule method.
Our results have allowed us to fix possible decay channels of these states
and found their widths. Investigations confirm that the doubly charged
diquark-antidiquarks are neither broad states nor very narrow resonances.

Observation of doubly charged tetraquarks may open new stage in exploration
of multiquark systems. In fact, resonances that are interpreted as hidden
charm (bottom) tetraquarks may be also considered as excited states of
charmonia (bottomonia) or their superpositions. The charged resonances can
not be explained by this way, and are serious candidates to genuine
tetraquarks. They may have diquark-antidiquark structure or be bound states
of conventional mesons. In the last case, charged and neutral conventional
mesons create shallow molecular states with large decay width. Therefore, it
is reasonable to assume that doubly charged tetraquarks presumably exist
only as diquark-antidiquarks, because binding of two mesons with the same
electric charge to form a molecular state due to repulsive forces between
them seems problematic.

The doubly charged tetraquarks deserve further detailed investigations.
These studies should embrace also $Z_{b\overline{c}}$-type states that
constitute a subclass of open charm-bottom states. Experimental exploration
and discovery of $Z_{c\overline{s}}$ and/or $Z_{b\overline{c}}$ tetraquarks
may have far-reaching consequences for hadron spectroscopy.

\section*{ACKNOWLEDGEMENTS}

The work of S.~S.~A. was supported by Grant No. EIF-Mob-8-2017-4(30)-17/01/1
of the Science Development Foundation under the President of the Azerbaijan
Republic. K.~A.~ thanks T\"{U}BITAK for the partial financial support
provided under Grant No. 115F183.

\end{document}